\documentclass[twocolumn]{aastex631}

\shorttitle{Coronal rain EUV spectroscopy} 

\shortauthors{Brooks et al.}
\journalinfo{To be published in The Astrophysical Journal}

\usepackage{grffile}

\begin{document}

\title{Spectroscopic Observations of Coronal Rain Formation and Evolution following an X2 Solar Flare}

\author[0000-0002-2189-9313]{David H. Brooks}
\altaffiliation{Present address: Hinode Group, ISAS/JAXA, 3-1-1 Yoshinodai, Chuo-ku, Sagamihara, Kanagawa 252-5210, Japan}
\affil{Department of Physics \& Astronomy, George Mason University, 4400 University Drive, Fairfax, VA 22030, USA}
\affil{University College London, Mullard Space Science Laboratory, Holmbury St. Mary, Dorking, Surrey, RH5 6NT, UK}

\author[0000-0003-4739-1152]{Jeffrey W. Reep}
\affil{Space Science Division, Naval Research Laboratory, Washington, DC 20375, USA}

\author[0000-0001-5503-0491]{Ignacio Ugarte-Urra}
\affil{Space Science Division, Naval Research Laboratory, Washington, DC 20375, USA}

\author[0000-0002-7983-3851]{John E. Unverferth}  
\affil{National Research Council Research Associate at the US Naval Research Laboratory, Washington, DC 20375, USA}

\author[0000-0001-6102-6851]{Harry P. Warren}
\affil{Space Science Division, Naval Research Laboratory, Washington, DC 20375, USA}

\begin{abstract}
A significant impediment to solving the coronal heating problem is that we currently only observe active region (AR) loops in their cooling phase.
Previous studies showed that the evolution of cooling loop densities and apex temperatures are insensitive to the magnitude, duration, and 
location of energy deposition. Still, potential clues to how energy is released are encoded in the cooling phase properties. The appearance
of coronal rain, one of the most spectacular phenomena of the cooling phase, occurs when plasma has cooled below 1\,MK, which sets constraints 
on the heating frequency, for example. Most observations of coronal rain have been made by imaging instruments. Here we report rare Hinode/EUV Imaging Spectrometer (EIS)
observations of a loop arcade where coronal rain forms following an X2.1 limb flare.
A bifurcation in plasma composition measurements between photospheric at 1.5\,MK and coronal at 3.5\,MK suggests that we are observing post-flare driven coronal rain.
Increases in non-thermal velocities and densities with decreasing temperature (2.7\,MK to 0.6\,MK) suggest that we are observing the formation 
and subsequent evolution of the condensations. Doppler velocity measurements imply that a 10\% correction of apparent flows in imaging data is reasonable.
Emission measure analysis at 0.7\,MK shows narrow temperature distributions, indicating 
coherent behaviour reminiscent of that observed in coronal loops. 
The space-time resolution limitations of EIS suggest that we are observing 
the largest features or rain showers. These observations provide insights into the heating rate, source, turbulence,
and collective behaviour of coronal rain from observations of the loop cooling phase. 
\end{abstract}

\section{Introduction}
Determining the mechanisms that heat the solar corona to high temperatures remains a difficult unsolved problem in solar physics.
There is a consensus that the coronal magnetic field is key, however, there is continuing debate on the precise details of how energy
is stored and dissipated. Two long standing lines of investigation involve the dissipation of Magnetohydrodynamic (MHD) waves and 
the reconnection of braided magnetic fields. In the MHD wave heating scenario, turbulent motions in the photosphere lead to the
propagation of Alfv\'{e}n waves along the magnetic field into the corona \citep{vanBallegooijen2011}. In the reconnection scenario, field line braiding leads to 
the build-up of magnetic stress later released by topological rearrangement \citep{Gold1960,Parker1983,Parker1988}. Other agents of energy transfer have been proposed in
recent years, such as chromospheric jets \citep{DePontieu2009,DePontieu2011} or solar tornadoes \citep{Wedemeyer2012}.
It seems clear that all these processes occur in the corona, 
and the challenge is to ascertain which one is dominant in a given structure at a specific time. 
Recent reviews are given by \cite{Klimchuk2006}, \cite{Parnell2012}, and \cite{Reale2014}.

One way to make progress on this problem is to compare state-of-the-art numerical simulations with the observations, but this has its
own difficulties. Current 3-D numerical simulations and observations cannot resolve the small spatial scales involved in the heating process \citep{Shay2001,Pontin2022}.
Radiation signatures of the process also appear after the mechanism has heated the plasma and we are already observing the cooling phase.
This is a significant barrier to further progress. 
Hydrodynamic simulations suggest that loop densities and apex temperatures do not depend clearly on the magnitude,
duration, or location of the energy deposition \citep{Winebarger2004}. This implies that observations need to be made early in the loop
evolution to infer information about the heating parameters. 
That being said, observations of the cooling phase can be a potential source of information on other aspects of the loop heating and cooling process, and therefore provide
useful constraints for numerical models.

One striking observational feature of cooling loops is the appearance of coronal rain \citep{Schrijver2001}.
Coronal rain clumps form in coronal loops as condensations due to the onset of a thermal instability when high radiative losses cause catastrophic cooling 
\citep{Muller2003}.
It can be created in numerical hydrodynamic simulations if the heating is concentrated at the loop footpoints 
\citep{Antiochos1999,Karpen2001,Muller2005}. 
This already tells us something about the heating mechanism, despite only observing the loops cooling, and sets limitations on the process itself
since, for example, Alfv\'{e}n waves generate uniform heating in some models \citep{Antolin2010}, but localize heating to the loop footpoints in others \citep{Downs2016}. 
Uniform (smooth) temperature distributions tend to inhibit the formation of coronal rain.

The heating frequency also impacts rain formation. High-frequency heating leads to Thermal Non-Equilibrium (TNE) and cyclic heating (evaporation) and cooling (condensation) behavior,
manifested as long-period intensity pulsations at high temperatures and coronal rain at lower temperatures \citep{Froment2015,Antolin2020}. There is some
evidence that the occurrence of TNE depends on a relatively narrow set of heating parameters \citep{Froment2018}. 
Physically, coronal rain can be produced in 3-D
radiative MHD simulations due to heating by the Ohmic dissipation of field line braiding as a result of convective motions in the photosphere \citep{Kohutova2020}.
This is impulsive heating, and coronal rain has also been observed in solar flares \citep{Jing2016} and non-flaring dynamic events \citep{Kohutova2019}.
The typical short duration non-thermal electron beam heating at loop footpoints in flares, however, does not seem to be able to produce coronal rain \citep{Reep2020},
though note that rain is generated if TNE and beam heating are combined. 

Observationally, most studies have focused on chromospheric data where the rain is most visible falling and tracing the paths of the magnetic field. Typical coronal
rain velocities are on the order of 30--100\,km s$^{-1}$ \citep{Antolin2012,Ahn2014}, with extreme values up to 200\,km s$^{-1}$ being reported \citep{Kleint2014}.
The widths of coronal rain clumps
are 200-300\,km on average with minima and maxima in the range of 150--800\,km for quiescent rain \citep{Antolin2012}, and 120\,km for
flare-driven rain \citep{Jing2016} when observed at high spatial resolution. There is a tendency
for the clumps to be wider at higher temperatures, and to show an association with the spatial resolution of the observing instrument \citep{Antolin2022a}.

The temperature evolution of the multi-thermal cooling plasma has been tracked using differential emission measure (DEM) techniques applied to SDO/AIA images
\citep{Antolin2015,Scullion2016,Kohutova2016,Froment2020}, and a fairly wide range of 
density measurements have been inferred indirectly: 
$\log (n/cm^{-3})$ = 10.3--11.4 from EUV absorption by hydrogen and helium \citep{Antolin2015},
$\log (n/cm^{-3})$ = 10.8--12.0 from an emission measure analysis of H$\alpha$ data \citep{Froment2020}, and
$\log (n/cm^{-3})$ = 11.9 from combining a volume estimate with a DEM measurement in flare-driven rain \citep{Scullion2016}.
For an extensive review of coronal rain and all these properties see \cite{Antolin2022a} and references therein.

There have been no measurements of elemental abundances, though coronal rain associated with impulsive events and flares would suggest the likelihood of a photospheric
composition \citep{Warren2014a,Warren2016}. 
In contrast, coronal rain forming in active region loops ought to show a coronal composition, thus making it easier to form due to the larger radiative losses.

Most of the information we know about coronal rain, and described here, comes from imaging data, chromospheric observations, and modeling, yet the rain features first condense and
form in the corona and cool through the transition region where EUV spectroscopic data is, in principal, obtainable by instruments such as the EUV Imaging 
Spectrometer \citep[EIS,][]{Culhane2007} on
Hinode \citep{Kosugi2007}. One difficulty with using EIS is that the transition region emission lines are weak. Another issue is that coronal rain is most readily
observed off-limb, and the main aim of Hinode is to combine coronal observations with measurements of the photospheric magnetic field, which necessarily leads
to an operational mode that favors on-disk observations and a tendency to make spatial maps.

Nevertheless, on 2023, March 3, spectacular coronal rain was observed following an X2.1 flare in AR 13234. EIS was observing off-limb in sit-and-stare mode,
positioned near the apex of a loop system where coronal rain was falling. This 
allowed us to obtain rare measurements of the formation and evolution of coronal rain. Here we report these new spectroscopic observations, including
density, temperature, and plasma composition measurements as the rain first condenses and forms. We also discuss the implications of our measurements.
\section{Data Processing}

\begin{figure*}
\centering
\includegraphics[viewport= 0 190 612 602,width=1.0\textwidth]{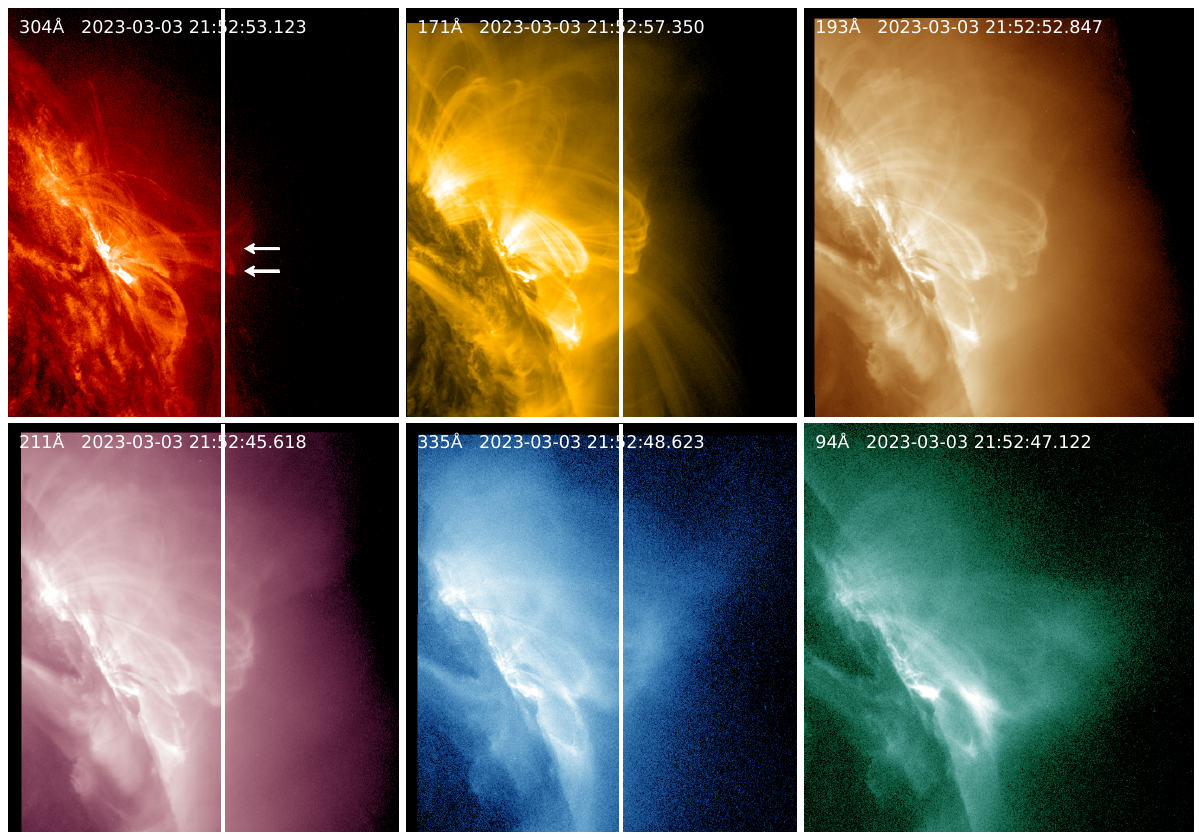}
\caption{ Context images from SDO/AIA showing the loop system in the west limb AR and the location of the EIS slit.
An animation showing the time-period of the flare and EIS observations is available in the HTML version of the article.
This figure is one frame from the animation, which lasts 12\,s and spans the period 16:01:21--22:58:45\,UT.
The filters and observation times are indicated in the legends. The filters are  sensitive to plasma temperatures
of 0.09\,MK (304\AA), 0.8\,MK (171\AA), 1.4\,MK (193\AA), 1.8\,MK (211\AA), 2.5\,MK (335\AA), and 7.1\,MK (94\AA) in active
regions, respectively, with contributions from continuum and higher temperature lines during flares.
We show the EIS slit position with the vertical white line on the AIA images used to coalign with EIS in Figure \ref{fig2}.
The arrows on the AIA 304\,\AA\, image show the locations that are tracked by the horizontal dashed lines in Figure \ref{fig2}.
}
\label{fig1}
\end{figure*}

We use data from the Solar Dynamics Observatory \cite[SDO,][]{Pesnell2012} Atmospheric Imaging Assembly \cite[AIA,][]{Lemen2012} 
in this work. AIA images the Sun at high spatial (0.6$''$ pixels) resolution and high time (12\,s) cadence in a range of filters sensitive
to different layers of the solar atmosphere. All of the data shown here were retrieved from the Stanford JSOC (Joint Science Operations 
Center) and processed with standard calibration procedures available in SolarSoftware \cite[SSW,][]{Freeland1998}.

Our main emphasis is on the spectroscopic measurements made by the Hinode/EIS.
EIS records solar spectra in two wavelength bands from 171--212\,\AA\, and 245--291\,\AA\, at moderate spatial resolution (1$''$ pixels) and
high spectral resolution (22\,m\AA). EIS typically observes in rastering or sit-and-stare mode. 
In raster mode, the fine mirror is moved to build up a series of slit exposures across an area of the Sun. In sit-and-stare mode, the mirror
does not move, and repeated exposures are made at the same position. The selected observing mode is a compromise between spatial coverage
and temporal cadence. Typically, sit-and-stare mode is used to observe the evolution of a particular feature at relatively high time cadence. For features
such as flares, there is an element of luck in being positioned in the right place at the right time. The data presented here are from a
relatively uncommon observation where the EIS slit was well positioned at the top of a limb AR loop arcade.

The observing sequence studied here used the 2$''$ slit in sit-and-stare mode, and took a sequence of 80 consecutive exposures. The exposure time for these observations was 40\,s, and the resulting cadence was 42\,s. An extensive
line-list was used, including spectral lines of \ion{He}{2}, \ion{O}{5}--\ion{O}{6}, \ion{Mg}{5}--\ion{Mg}{7}, \ion{Si}{7}, \ion{Si}{10}, 
\ion{Al}{9}, \ion{S}{10}, \ion{S}{13}, \ion{Ar}{14}, \ion{Ca}{14}--\ion{Ca}{17}, \ion{Fe}{8}--\ion{Fe}{17}, \ion{Fe}{23}, and \ion{Fe}{24}. This selection
includes several density and elemental abundance diagnostics, and a wide range of lines suitable for measuring the atmospheric temperature structure 
(differential emission measure). We discuss the specific diagnostics we use in Sect. \ref{methods}.

The observing sequence was run 4 times on 2023, March 03, from 18:57:34--22:41:16\,UT. We processed the level-0 FITS files using the standard eis\_prep
procedure in SSW. This code removes the dark current pedestal, handles warm, hot, dusty pixels and cosmic ray strikes, and converts
the data to physical units by applying the radiometric calibration. For this analysis, we used the absolute calibration of \cite{Warren2014}.

\begin{figure*}
\centering
\includegraphics[viewport= 0 170 612 622,width=1.0\textwidth]{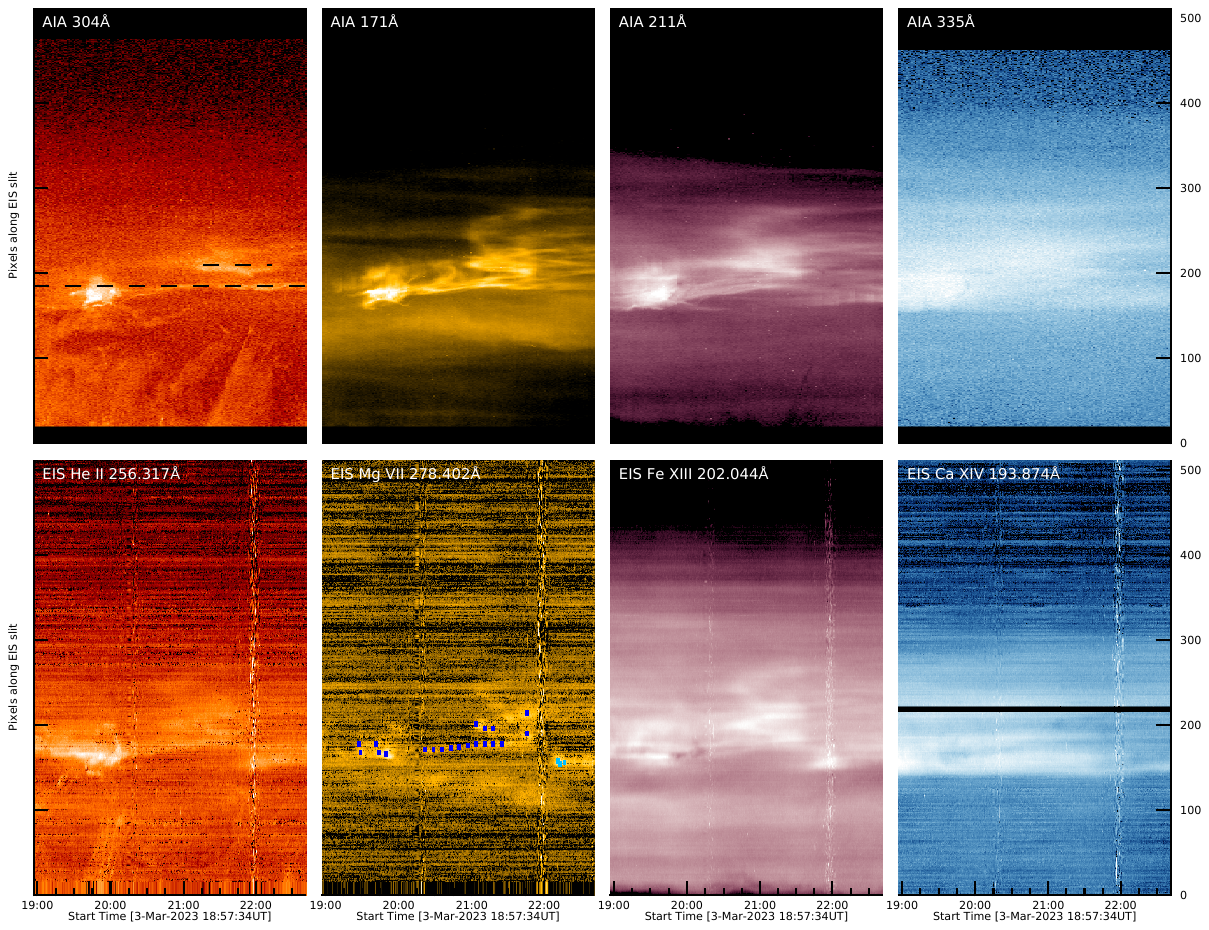}
\caption{ Time-distance plots showing the accuracy of the AIA/EIS coalignment and identifying the features associated
with coronal rain and the loop system. Top row: AIA time-distance plots at the position of the EIS slit. Bottom row:
EIS time-distance plots. We show the AIA filters and EIS spectral lines used in the legends. The dashed lines on the
AIA 304\AA\, image indicate trails produced by the continuous passage of rain features across the EIS slit. The Y-pixel
positions corresponding to these trails are highlighted by arrows in the animation for Figure \ref{fig1}. The dark blue boxes show
the positions where the spectroscopic diagnostic measurements were made. The light blue boxes show the positions where extra
analysis was performed on the Mg data (see text). The horizontal black line in the \ion{Ca}{14} 193.784\,\AA\, image is a result
of contamination on the detector.
}
\label{fig2}
\end{figure*}

The EIS team routinely monitor the pointing relative to AIA full-disk images using wide-slit (266$''$) data. This allows calibration of the EIS-AIA
offset. We coaligned the EIS and AIA data taking this offset into account. We also corrected the EIS pointing for the satellite jitter in the X- and Y-directions. 
\section{Data Analysis and Observational Results}
\label{methods}

\subsection{Overview of activity}
\label{overview}

Active region NOAA 13234 crossed the solar disk from 2023, February 19 until March 5. Originally classified as a bipolar $\beta$ sunspot group, it
developed a $\beta \gamma \delta$ configuration with continuous flux emergence in the core, and rotational behavior producing shearing around the 
mixed-polarity trailing $\delta$ spot. It was the target of a Max Millenium Major Flare Watch for long periods, and produced multiple flares greater
than M-class. An X2.1 event occurred as the region sat on the west limb, peaking at 17:52\,UT\, on 2023, March 3 (SOL2023-03-03) according to 
the Hinode Flare Catalogue \citep{Watanabe2012}.
The flare itself had a fast rise phase but fairly slow decay. EIS started observing about an hour later, and we observed
the coronal rain with the sit-and-stare flare mode sequence.

\begin{figure*}
\centering
\includegraphics[viewport= 0 170 612 622,width=1.0\textwidth]{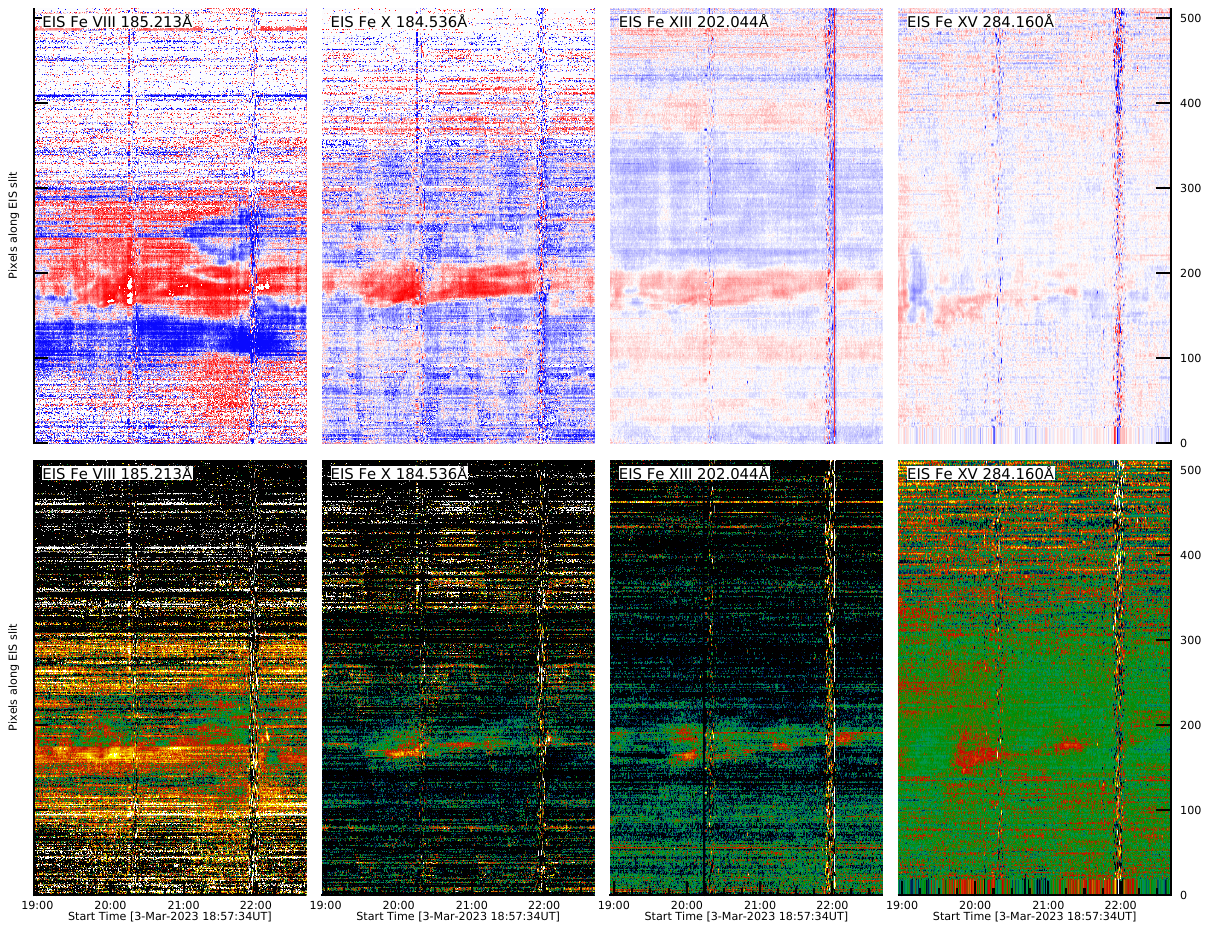}
\caption{ Time-distance plots of Doppler (top row) and non-thermal (bottom row) velocities derived from the EIS data.
The spectral lines are indicated in the legends and cover a broad range of theoretical temperatures: 
0.5\,MK (\ion{Fe}{8} 185.213\,\AA), 1.0\,MK (\ion{Fe}{10} 184.536\,\AA), 1.7\,MK (\ion{Fe}{13} 202.044\,\AA), and 2.2\,MK (\ion{Fe}{15} 284.160\,\AA).
Blue/red indicates emission moving towards/away from the observer. The Doppler velocity maps are scaled to $\pm$ 25\,km s$^{1}$.
The non-thermal velocity maps are scaled from 1--80\,km s$^{-1}$ (\ion{Fe}{8} 185.213\,\AA\, and \ion{Fe}{10} 184.536\,\AA) or 1--50\,km s$^{-1}$
(\ion{Fe}{13} 202.044\,\AA\, and \ion{Fe}{15} 284.160\,\AA).
}
\label{fig3}
\end{figure*}

Figure \ref{fig1} shows multiple AIA images of AR 13234 and the coaligned position of the EIS slit at 21:52:45--21:52:57\,UT. The figure is one frame from
the animation available in the online manuscript. The animation shows the same image format, and overplots the position of the EIS slit when the sit-and-stare
observations are running. Prior to the flare onset, we see quiescent coronal rain falling around the active region in the 304\,\AA\, images. Bright
condensations appear to form at the loop tops, and these are particularly clear from 16:42\,UT. They are seen as falling dark absorption features
that are visible in all the other filters except 94\,\AA. 
The flare starts at 17:42\,UT \citep{Watanabe2012}, and is quickly followed by a bright eruption surrounding another dark absorption feature
in the 304\,\AA\, images. This feature appears to be an erupting filament, and is also visible in all filters except 94\,\AA. The erupting loop arcade
is very prominent in the animation. This eruption destabilizes the loop arcade, and appears to energize the loop top region where the coronal rain 
had previously been forming. The system then begins to relax, and similar bright coronal rain (in 304\,\AA) and dark absorption features (other wavelengths except 94\,\AA)
start to form and fall again. To be clear, quiescent active region coronal rain is falling throughout the period of the flare. There is no break in rain activity due to the flare, though
the loops hosting the rain are disrupted by it and temporarily obscured in the animation by the bright flare emission. As the brightness subsides these loops are visible again and rain 
is falling. 

As EIS begins to observe at 18:57:34\,UT, 65\,mins after the flare peak, the slit is well positioned close to this coronal rain formation region at the loop tops. 
It is unclear, however, whether the post-flare rain EIS observes is actually driven by the flare, or whether it is quiescent rain typical of this active region that has 
restarted once the disruption of the flare is over i.e. rain 
similar to that which was observed by AIA prior to the flare. This is an important question since these observations could provide new constraints for numerical modeling. 
Current models fail to adequately reproduce flare-driven rain, since the heating in a flare is so strong that conduction quickly diffuses away
any perturbation that can give rise to the growth of a rain clump \citep{Reep2020}. 
The only way we confidently know how to produce rain in loop simulations is through thermal non-equilibrium with steady footpoint heating,
which seems contradictory to the impulsive flare scenario. Furthermore, there are significant differences in the modeling if the rain forms close to the flare impulsive
phase, rather than well into the gradual phase. In our observations, 
if the rain is flare driven, it is immediately visible close to the impulsive phase. In section \ref{pcil} we use EIS plasma composition measurements to 
conclude that this rain is indeed flare-driven. 

Two positions are pointed out with arrows
in the animation of Figure \ref{fig1}. These correspond to the locations of bright tracks seen in the time-distance plots of Figure \ref{fig2}.
Figure \ref{fig2} shows time-distance plots for several AIA filters and EIS spectral lines that correspond to broadly similar formation temperatures.
These demonstrate the accuracy of the EIS-AIA coalignment and highlight the features produced by the coronal rain. Looking at the AIA 304\,\AA\, and 
EIS \ion{He}{2} 256.317\,\AA\, plots we can see a bright feature around Y-pixels 150--200 and 19:30--20:00\,UT. This corresponds to the bright loop
top blobs of coronal rain that form and cross the EIS slit at the position of the lower arrow in the animation. Rain is crossing here almost throughout
the EIS observing period and it produces the horizontal track in the 304\,\AA\, time-distance plot (highlighted with the lower dashed-line in Figure \ref{fig2}).
Slightly higher and later, around Y-pixel 210 and 21:30--22:00\,UT, there is another horizontal trail (highlighted with the upper dashed-line). This
corresponds to the weaker material forming and crossing at the position of the upper arrow in the animation. That coronal rain is visible in EIS transition region
lines (\ion{Mg}{6}) has been demonstrated before \citep{Ugarte2009}, and the coronal rain trails here are clearly visible in the EIS \ion{He}{2} 256.317\,\AA\,
and \ion{Mg}{7} 278.402\,\AA\,
images, though the morphology is not exactly the same due to the different temperature and instrument sensitivities, the orbital motion of the EIS slit, and the accuracy of the coalignment.
The trails in the time-distance plots for the other wavelengths (\ion{Fe}{13} 202.044\,\AA\, and \ion{Ca}{14} 193.874\,\AA) indicate that the hotter host loops are being observed at these positions at higher temperatures.

\begin{figure*}
\centering
\includegraphics[viewport= 120 180 542 612,width=0.45\textwidth]{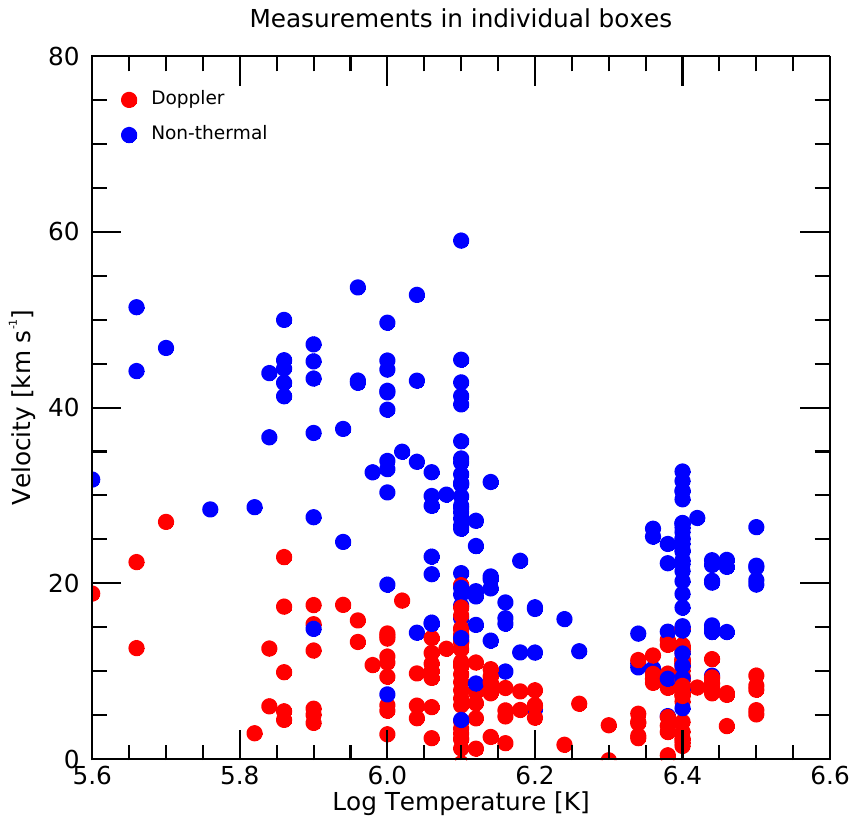}
\includegraphics[viewport= 120 180 542 612,width=0.45\textwidth]{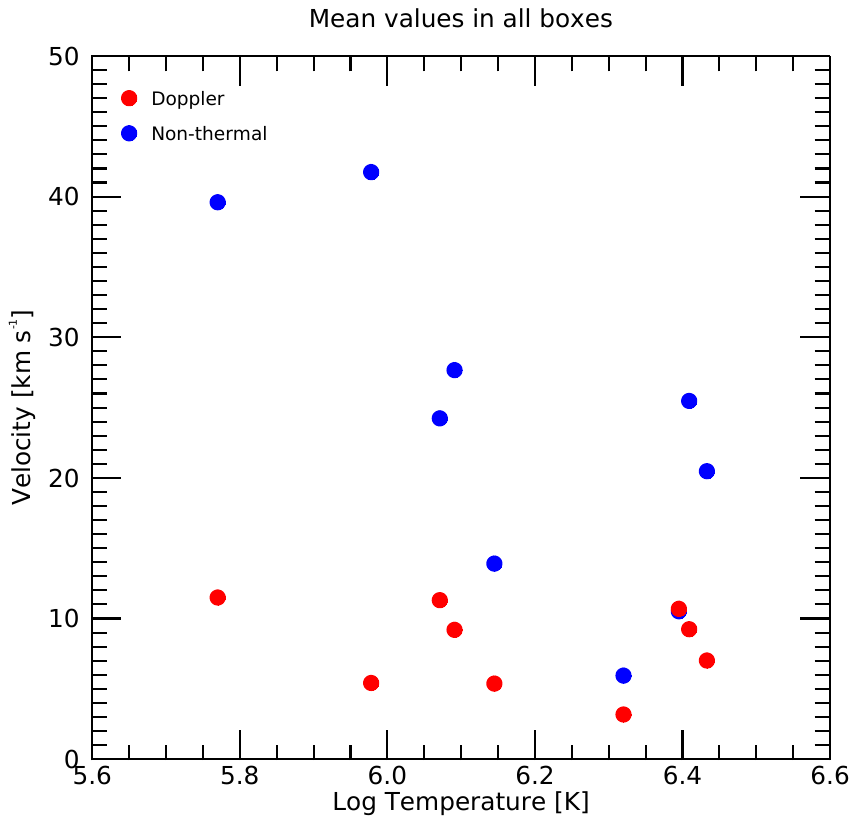}
\caption{ Coronal rain and cooling host loop velocity analysis. 
Left panel: Doppler (red) and non-thermal (blue) velocities measured in the dark blue boxes of Figure \ref{fig2} as a function of effective formation temperature.
Lines from \ion{Fe}{8}--\ion{Fe}{16} were used. Right panel: Mean values of the Doppler (red) and non-thermal (blue) velocities in all the boxes, plotted
as a function of the average effective formation temperature.
}
\label{fig4}
\end{figure*}

To measure the spectroscopic properties of coronal rain in EIS sit-and-stare observations, we need to ensure sufficient spatio-temporal averaging
to obtain a good signal-to-noise (S/N) ratio in the spectrum. Previous studies give guidance on how to approach this problem. 
Given the range of widths and velocities quoted in the introduction, a large (300\,km) and slow (30\,km s$^{-1}$) clump is still smaller than the 
725\,km spatial pixels of EIS, and would pass through the 2$''$ slit location on the order of the exposure time (40\,s). It means that in most cases the coronal
rain clumps are contained within 1 EIS spatio-temporal pixel, and, unfortunately, this implies that we are not measuring individual clumps in our EIS observations.

Nevertheless, previous observations also show that coronal rain clumps 
appear to evolve coherently on larger spatial scales of a few Mm \citep{Sahin2022}, and this has been referred to as a coronal rain shower. 
It therefore suggests that although we are likely not measuring the properties of individual clumps of coronal rain, we are making measurements of 
the largest clumps or the rain showers,
or in fact averages of clumps that are representative of the whole shower or loop envelope.
We therefore averaged over the smallest areas possible, consistent with obtaining a good S/N ratio for the spectra. 
This resulted in box sizes of 3$'' \times$5$''$ ($\sim$2.2$\times$3.6\,Mm). 

The value of the S/N ratio within these boxes, defined here as the ratio of the peak intensity to the background intensity, that is adequate to obtain a good fit to the observed profile depends on the specific spectral line.
For the strong \ion{Fe}{8}--\ion{Fe}{16} lines, the values fall in the range of 5.8 to 83.5; with the lowest value for \ion{Fe}{9} 197.862\,\AA\, and the highest value for \ion{Fe}{12} 195.119\,\AA. 
We initially selected 20 boxes in total in the coronal rain trails of the \ion{Mg}{7} 278.402\,\AA\, time-distance plot (shown in dark blue in Figure \ref{fig2}). 
Although the coronal rain is seen most clearly
in \ion{He}{2} 256.317\,\AA, all of the EIS diagnostics are sensitive to higher temperatures. In particular, the \ion{Mg}{7} 280.737/278.402 diagnostic ratio
allows us to measure the density in the coronal rain itself, so we needed to ensure that we chose areas where there is relatively strong emission in these lines. 
This is why we used \ion{Mg}{7} to identify locations. The other density and composition
measurements are made at higher temperatures in the host loops where the rain forms as the loop cools.
We confirmed that the boxes we chose were large enough to get a good signal in all the lines analyzed, sometimes with the exception of the 
weak density diagnostic \ion{Mg}{7} 280.737\,\AA\, line. This is discussed further in Section \ref{mgvii}. Following the conclusions from that section, we selected 3 separate boxes 
(light blue in Figure \ref{fig2}) to supplement the analysis.

\subsection{Velocities in the host coronal loops}
\label{velocities}
Figure \ref{fig3} shows time-distance plots of Doppler and non-thermal velocities derived from several EIS spectral lines spanning the range of 
temperatures we analysed. Since EIS does not have an absolutely calibrated wavelength scale, these maps show relative Doppler velocities.
As with the intensity maps in Figure \ref{fig2}, the spatial offsets between the long- and short-wavelength detectors were corrected using
the neural network model of \cite{Kamio2010}. The same model was used to adjust for the orbital motion of the spectrum across the detector.
This model uses an artificial neural network to establish a relationship between the drift of the spectrum across the detector and
temperature variations inside the instrument due to the changing thermal environment around the satellite orbit. The resultant empirical
model reproduces the spectral drift to within 4.4\,km s$^{-1}$, and also returns the spectral curvature and spatial offsets.
The velocities were also calculated with respect to a reference wavelength where the line-of-sight flows are assumed to average to zero.
Typically, for on disk observations, a region of quiet Sun is used, but even there coronal lines are expected to be slightly blue-shifted \citep{Peter2001}.
Here, we are observing off-limb, so these blue-shifts are in the radial direction and the average blue-shift along the line-of-sight is expected
to be smaller leading to a lower uncertainty and more accurate velocity measurements \citep{Warren2011}. We chose the top 100 pixels of the slit across
the whole time-sequence and averaged all the spectra to produce reference wavelengths for each line. The final velocity maps were detrended for
residual variations as described in the appendix of \cite{Brooks2020}. The non-thermal velocities were calculated following the methods of
\cite{Brooks2016}. Briefly, the absolute intensity calibration has a tendency to increase the line widths, so we used the uncalibrated
data to calculate the non-thermal velocities. The line widths were also corrected for the instrumental width variation N-S along the 
slit using the routine eis\_slit\_width prepared by Young (2011)
\footnote{EIS Software Note No.7, SolarSoftware, \$SSW/Hinode/eis/doc/eis\_notes} and available in SSW.

\ion{Fe}{8} 185.213\,\AA\, is formed closest in temperature to the \ion{Mg}{7} 278.402\,\AA\, line where we observe the coronal rain. We see
a trail of red-shifted plasma in \ion{Fe}{8} 185.213\,\AA\, (top left panel) where we see a rain related intensity trail in 
\ion{Mg}{7} 278.402\,\AA\, (Figure \ref{fig2}), however, the red-shifts are not co-spatial and cover a more extensive area.
The trail of increased non-thermal velocity in \ion{Fe}{8} 185.213\,\AA\, also appears to correspond better with the
intensity trail in \ion{Mg}{7} 278.402\,\AA. This suggests that the Doppler velocity measurements may be picking up emission from the 
surrounding loops, or that the extent of the coronal rain clumps within the host loop increases as the loop cools.
At higher temperatures, we see narrower and weaker red-shift trails, and clumps of enhanced
non-thermal velocities. Our actual measurements of the rain and host loop Doppler and non-thermal velocities are, of course, made in the 
selected boxes.

Figure \ref{fig4} shows the Doppler and non-thermal velocity measurements obtained from the \ion{Fe}{8}--\ion{Fe}{16} lines
for the dark blue boxes in Figure \ref{fig2}. To be clear, there are two velocity measurements (Doppler and non-thermal) for each spectral
line in each of the 20 boxes. In the left panel, showing all the measurements, the non-thermal velocities increase with decreasing temperature, but there is no obvious
trend with temperature in the Doppler velocities. This plot shows a lot of scatter, however, 
the trends are evident and more clearly seen in the summary plot that shows
the average value at each temperature in all the dark blue boxes (right panel).
We plot the unsigned Doppler velocities since we are not concerned with the direction of the flows along the line-of-sight, but in
any case 90\% of the measurements are red-shifted. They fall in the range of 0.1--27\,km s$^{-1}$ with a mean value of
8.6\,km s$^{-1}$. This is consistent with the mean Doppler velocity of $\sim$10\,km s$^{-1}$ (also red-shifted) obtained by \cite{Antolin2012}
from spectropolarimetric data \citep[CRISP,][]{Scharmer2008} in the chromospheric H$\alpha$ line. \cite{Antolin2012} also found a mean plane-of-sky
velocity of $\sim$65\,km s$^{-1}$, suggesting that a 10\% correction for EIS Doppler velocities up to $\sim$30\,km s$^{-1}$ is reasonable.
There is no clear trend with temperature. 
According to previous work, rain clumps show moderate velocities (30--50\,km s$^{-1}$) at the loop top, accelerating to 100--150\,km s$^{-1}$ 
further down the loop leg \citep{Oliver2014}. Our new spectroscopic measurements are smaller at the loop top, and the rain clumps would likely 
need to cool to chromospheric temperatures before they are accelerated significantly. This suggests
that we are observing the rain clumps too early in the formation process to detect any evidence of acceleration
while cooling e.g. due to gravity.

The non-thermal velocities fall in the range of 0--59\,km s$^{-1}$ with
a mean value of 23.3\,km s$^{-1}$. The trend of increasing non-thermal velocity with decreasing temperature shows a strong linear
Pearson correlation coefficient of -0.6. There are several possible sources of this non-thermal broadening and assessing their relative contributions
benefits from measurements of the electron density. We investigate the densities in the following sections and discuss the implications for the broadening
further in Section \ref{broadening}.

\begin{figure*}
\centering
\includegraphics[viewport= 120 180 542 612,width=0.45\textwidth]{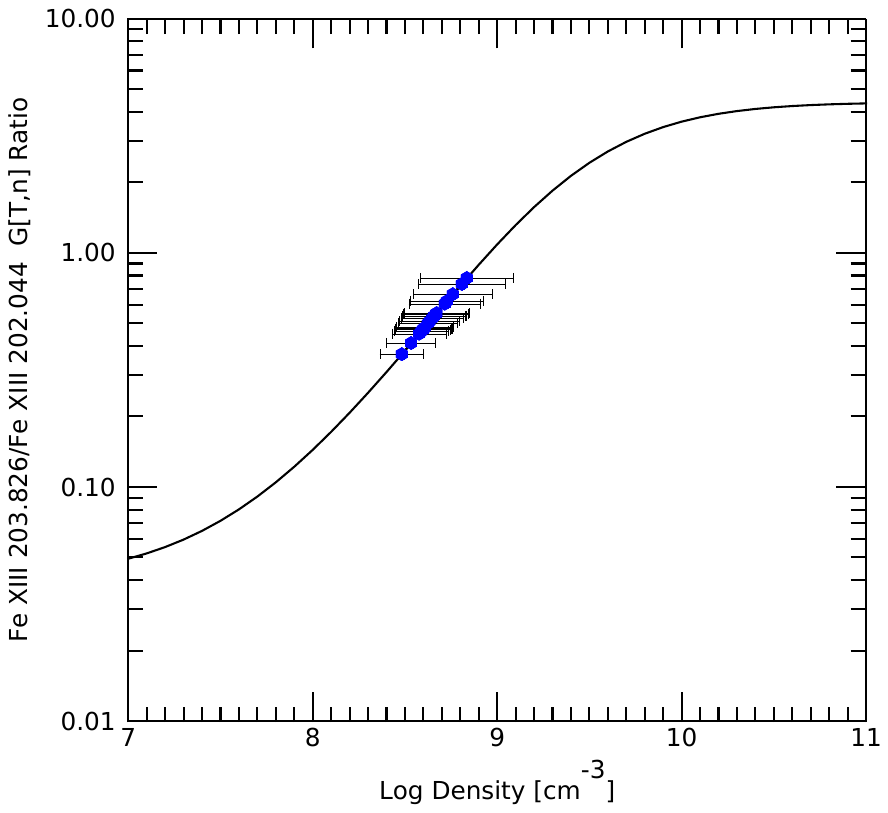}
\includegraphics[width=0.45\textwidth]{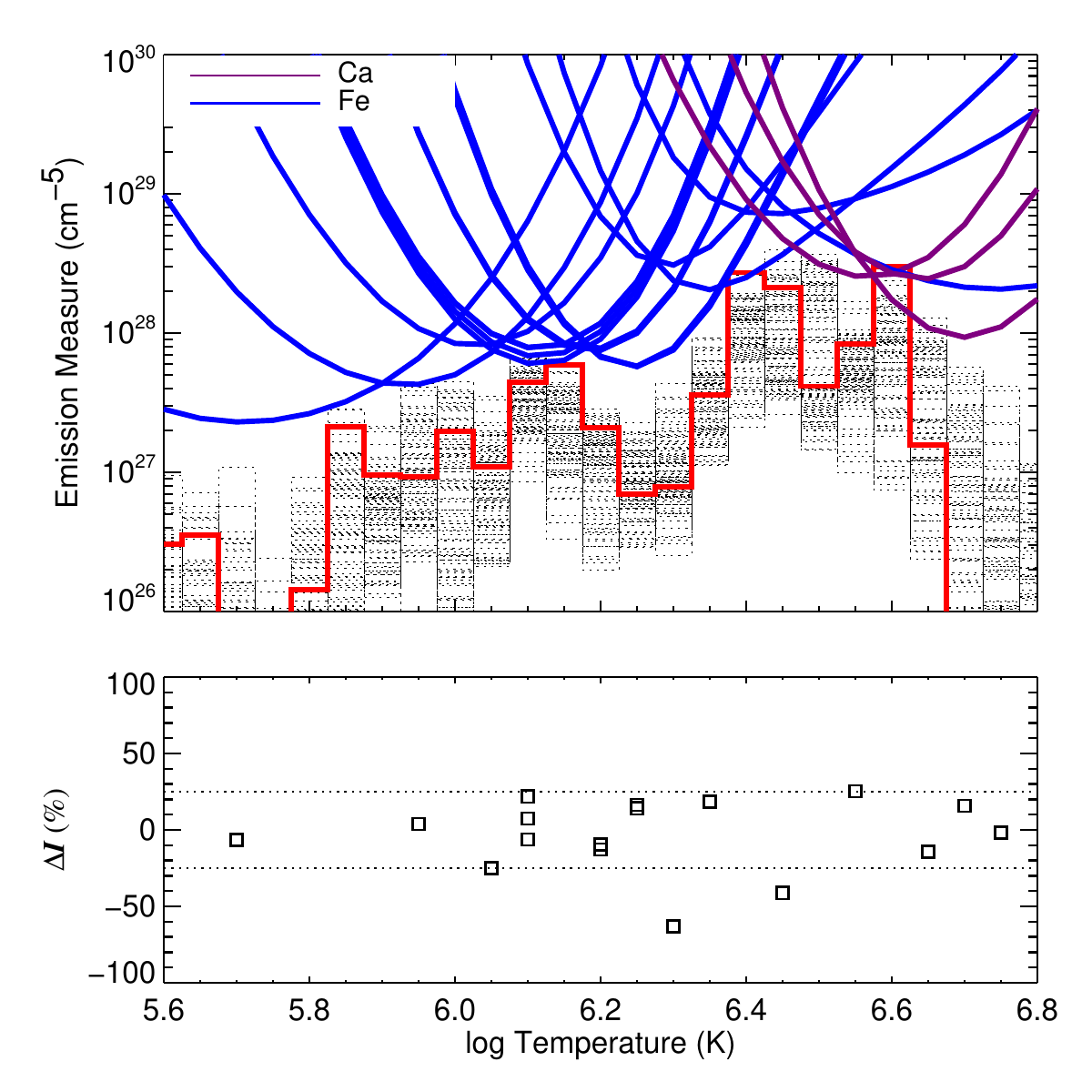}
\caption{ Cooling host loop coronal density and temperature analysis. Left panel: logarithmic electron density derived from the theoretical
\ion{Fe}{13} 203.826/202.044 line ratio. The ratio is shown by the solid black line. The blue dots indicate the measured
densities within the dark blue boxes shown in Figure \ref{fig2}.
The error bars show the range in densities that result from uncertainties in the 
observed ratio due to the radiometric calibration. 
Right panel (top): the emission measure (EM) distribution calculated
from the Fe and Ca lines. The blue and magenta lines are EM loci curves that indicate the upper limits to the emission 
measure. The solid boxcar red line is the best fit solution from the MCMC algorithm. The dotted black lines show different 
realizations of the Monte Carlo solutions (see text). 
Right panel (bottom): differences between the observed and calculated
intensities (expressed as a percentage) as a function of temperature. The dotted horizontal lines indicate an agreement
level of 25\%.
}
\label{fig5}
\end{figure*}

Inspection of the values within the region covered by the boxes did not reveal any correlation, or obvious temporal delay, between the unsigned Doppler and 
non-thermal velocities below 1\,MK, where we are potentially making measurements in the rain rather than in the host loops. 
Note that we calculated the non-thermal velocities assuming that the ion temperature is equal to the effective formation temperature i.e. the 
temperature of the peak of the contribution function convolved with the DEM distribution of the box. We also
used that temperature to plot the values.

\begin{deluxetable}{lccc}
\tabletypesize{\small}
\tablecaption{EIS DEM analysis on 03-Mar-2023}
\tablehead{
\multicolumn{1}{c}{ID} &
\multicolumn{1}{c}{I$_{obs}$} &
\multicolumn{1}{c}{I$_{calc}$} &
\multicolumn{1}{c}{$\Delta [\%]$} 
}
\startdata
Fe VIII 185.213 & 451.4$\pm$99.6 & 475.6 & 5.4 \\
Fe IX 197.862 & 99.6$\pm$22.1 & 104.8 & 5.2 \\
Fe X 184.536 & 846.6$\pm$186.7 & 727.0 & -14.1 \\
Fe XI 192.813 & 356.2$\pm$78.6 & 322.9 & -9.3 \\
Fe XI 180.401 & 3502.6$\pm$771.7 & 3166.3 & -9.6 \\
Fe XI 188.216 & 1304.3$\pm$287.2 & 1536.7 & 17.8 \\
Fe XI 188.299 & 899.3$\pm$198.0 & 932.3 & 3.7 \\
Fe XII 195.119 & 2302.6$\pm$506.7 & 1817.7 & -21.1 \\
Fe XII 192.394 & 712.8$\pm$156.9 & 583.0 & -18.2 \\
Fe XIII 203.826 & 750.5$\pm$167.2 & 875.6 & 16.7 \\
Fe XIII 202.044 & 1127.9$\pm$248.6 & 1345.8 & 19.3 \\
Fe XIV 264.787 & 2566.7$\pm$564.7 & 935.7 & -63.5 \\
Fe XV 284.160 & 10810.2$\pm$2378.3 & 12182.4 & 12.7 \\
Fe XVI 262.984 & 1469.6$\pm$323.4 & 855.8 & -41.8 \\
Fe XVII 254.87 & 30.9$\pm$7.6 & 31.9 & 3.2 \\
Ca XIV 193.874 & 97.4$\pm$21.6 & 118.8 & 22.1 \\
Ca XV 200.972 & 76.8$\pm$17.7 & 60.2 & -21.5 \\
Ca XVI 208.604 & 24.3$\pm$16.9 & 29.6 & 22.1   
\enddata
\tablenotetext{}{EIS line intensities are in units of erg cm$^{-2}$ s$^{-1}$ steradian$^{-1}$.}
\label{table1}
\end{deluxetable}

\subsection{Density and temperature analysis of the host loops}
\label{dem}
To compute the DEM within the dark blue boxes we used the Markov-Chain Monte Carlo (MCMC) code available in the PintOfAle software package 
\citep{Kashyap1998,Kashyap2000}, together with the CHIANTI database \citep{Dere1997} v.10 \citep{DelZanna2021} for all the atomic data.
We also assume that the elemental abundances correspond to those of the solar photosphere \citep{Scott2015a,Scott2015b}.
The MCMC algorithm solves the inversion equation
\begin{equation}
I_{ij} = \int G(T,n) \phi(T) dT
\label{eq1}
\end{equation}
by performing a series of reconstructions (100 in our case) that reduce the differences between the calculated and measured spectral line intensities.
In Equation \ref{eq1}, $I_{ij}$ is the line intensity for the atomic transition between levels $i$ and $j$, $\phi(T)$ is the DEM as a function of temperature, $T$, and $G(T,n)$ is the usual contribution
function dependent on temperature and density, $n$. To reduce the problem to a temperature inversion, we also measure the electron density using
the \ion{Fe}{13} 202.044/203.826 diagnostic ratio. 

The density is a key quantity of interest. For the \ion{Fe}{13} ratio, we are making a measurement around 1.7\,MK, which is close to the expected
peak temperature of `warm' active region loops \citep{Reale2014}. Such loops are observed in their cooling phase \citep{Ugarte2009}, and coronal 
rain forms within them if they reach a low enough temperature. Since the observations take place after the X2.1 flare, it is unclear whether we
are observing loops cooling in the active region after the influence of the flare has ended, or actual post-flare loops. In either case, 
at these temperatures, we are measuring the densities in the host loop where the coronal rain later forms.

Warm loops have narrow temperature distributions \citep{Warren2008}, but it is clear that in these observations we are observing the emission from 
multiple different structures at different temperatures along the line-of-sight (coronal rain, warm loops, high-temperature loops). Initially,
we therefore derive the DEM including all the spectral lines in our analysis. This is because we use this DEM primarily for measuring the
elemental abundances. In section \ref{mgvii}, we attempt to isolate the emission from the coronal rain itself.

Figure \ref{fig5} (left panel) shows the theoretical \ion{Fe}{13} 203.826/202.044 intensity ratio plotted against density, and the measurements in
all the dark blue boxes as blue dots. Densities in the host loops fall in the range of $\log (n/cm^{-3})$ = 8.5--8.8, with a mean value of $\log (n/cm^{-3})$ = 8.65.
We show an example of the emission measure (EM) distribution from one of the boxes in the right panel of Figure \ref{fig5}. The distribution covers a broad range of temperatures
reflecting the fact that we are observing emission from the transition region and active region loops, all the way up to the high temperature active region
or post-flare loops. The EM in this case peaks around 2.8\,MK, and is well constrained by the high temperature Ca lines. A well-constrained EM at these
temperatures is critical for the plasma composition measurements. The specific lines used, together with
the observed and calculated intensities, and the percentage differences between them are shown in Table \ref{table1}. 
The figure and table show that 16/18 (89\%) of the line intensities are reproduced to within the EIS intensity
calibration uncertainty of $\sim$23\% \citep{lang2006}.

\begin{figure}
\centering
\includegraphics[viewport= 120 180 542 612,width=0.45\textwidth]{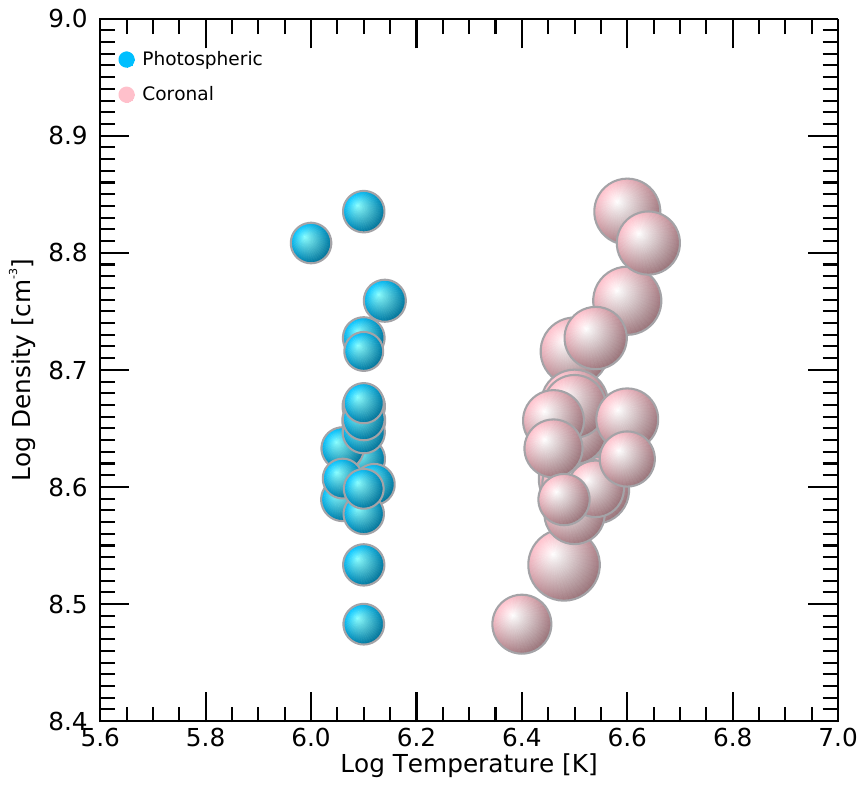}
\caption{ Cooling host loop coronal composition analysis. Bubble plot of the FIP bias factors measured in the dark blue boxes of Figure \ref{fig2} using
the Si/S and Ca/Ar abundance diagnostic ratios. 
The results are plotted at the electron density measured using \ion{Fe}{13} and the effective formation temperature of \ion{S}{10} 258.375\,\AA\, ($\log T \sim$6.1) and \ion{Ca}{14} 193.874\,\AA\, ($\log T \sim$6.5).
The size of the bubble is proportional to the magnitude of the FIP bias. Sky blue indicates that the FIP bias is photospheric ($<$1.5), and pink indicates that
it is coronal ($>$1.5). The bifurcation at the two temperatures is evident.
}
\label{fig6}
\end{figure}

\subsection{Plasma composition in the host loops}
\label{pcil}
To measure the plasma composition we compute the FIP bias (ratio of coronal to photospheric composition) by modeling two intensity ratios 
that are sensitive to the elemental abundances: \ion{Si}{10} 258.375\,\AA/\ion{S}
{10} 264.223\,\AA, and \ion{Ca}{14} 193.784\,\AA/\ion{Ar}{14} 194.396\,\AA. The DEM is derived in each box using the method described in Section \ref{dem}.
Only lines from the low-FIP elements Fe and Ca are used in this analysis (see Table \ref{table1}). The Ca lines are necessary to provide constraints on
the DEM at the high temperatures where the Ca/Ar ratio is determined. Their impact on the DEM at the formation temperature of the Si/S ratio is less
important. As discussed in Section \ref{dem}, we assumed photospheric abundances when calculating the DEM. From here, the method to calculate the Si/S
ratio is the same as we have used previously \citep{Brooks2011,Brooks2015}. The DEM solution is adjusted to match the \ion{Si}{10} 258.375\,\AA\,
intensity. This takes account of potential cross-detector calibration issues resulting from the fact that most of the Fe lines used in the DEM analysis lie on the short-wavelength detector, whereas the Si and S
lines lie on the long-wavelength detector. The \ion{S}{10} 264.223\,\AA\, line intensity is then computed, and the ratio of the calculated to observed
intensity is the FIP bias, $f_{Si/S}$. For the Ca/Ar ratio, no adjustment to the DEM is made because both the lines lie on the short-wavelength detector, and the 
DEM is already constrained by the Ca lines. The ratio of the predicted and observed intensities of \ion{Ar}{14} 194.396\,\AA\, yields the FIP bias, $f_{Ca/Ar}$.
This method attempts to remove any diagnostic dependence on temperature and/or density. The Ca/Ar ratio has been used successfully for previous 
abundance measurements in flares \citep{Doschek2015,Baker2019}.

Figure \ref{fig6} is a bubble plot showing the measured values of $f_{Si/S}$ and $f_{Ca/Ar}$ as a function of the electron density and effective formation
temperature in each box. There is a clear difference in the distributions at the two formation temperatures. For Si/S ($\log T \sim$6.1) the values
fall in the range of $f_{Si/S}$ = 1.2--1.5, with a mean of $f_{Si/S}$ = 1.3, indicating photospheric abundances. 
For Ca/Ar ($\log T \sim$6.5) the values fall in the range of $f_{Ca/Ar}$ = 2.0--4.0, with a mean
of $f_{Ca/Ar}$ = 3.1, indicating coronal abundances.

Generally we have found that the scaling factor between Si and Fe can be up to a factor of 2. This contrasts with measurements earlier in the mission
where the scaling was always less than the calibration uncertainty \citep[see, e.g.,][]{Brooks2011}. It suggests there has been 
a change in the short- to long-wavelength
calibration and this is currently being assessed \citep{DelZanna2023}. Our method corrects for any calibration evolution.

\begin{figure*}
  \centerline{%
    \includegraphics[viewport= 20 330 612 462,width=1.0\textwidth]{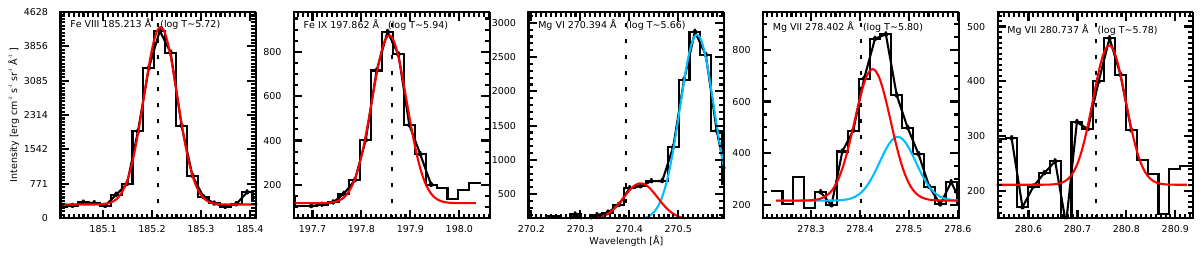}} %
  \centerline{%
    \includegraphics[viewport= 20 330 612 462,width=1.0\textwidth]{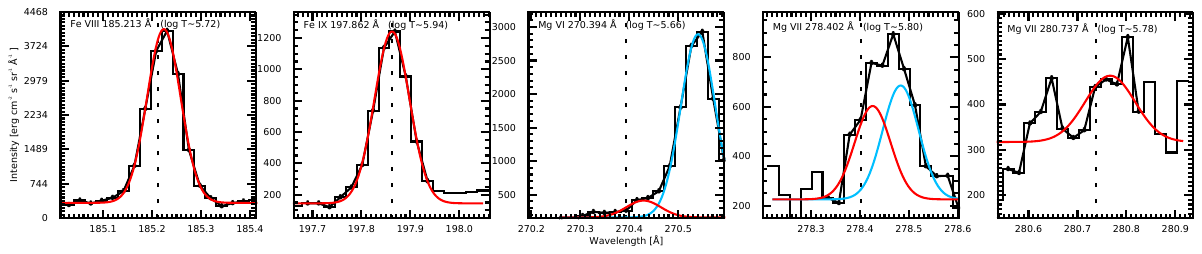}} %
  \caption{Example line profiles in the coronal rain showers.
Top row: example box (from Figure \ref{fig2}) where the weakest line (\ion{Mg}{7} 280.737\,\AA) is still strong enough that we consider it well-fitted.
The black histograms and small stars show the EIS data. The red lines show the Gaussian fits to the Fe and Mg lines of interest. 
The light blue lines show Gaussian fits to the blends i.e. in the 3rd column, \ion{Fe}{14} 270.519\,\AA\, (light blue) blends with \ion{Mg}{7} 280.737\,\AA\, (red), and in the 4th column,
\ion{Si}{7} 278.445\,\AA\, (light blue) blends with \ion{Mg}{7} 278.402\,\AA\, (red). The solid black curves show the composite fit from the main and blended lines.
The Fe and Mg spectral lines are indicated in the legends along with their theoretical formation temperatures. The dotted vertical lines show the rest wavelength.
\ion{Mg}{7} 278.402\,\AA\, and \ion{Mg}{7} 280.737\,\AA\, are the density diagnostic pair,
and \ion{Mg}{6} 276.153\,\AA\, is the other Mg line used in the coronal rain emission measure analysis.
\ion{Fe}{8} 185.213\,\AA\, and \ion{Fe}{9} 197.862\.\AA\, are the weakest Fe lines used in our analysis, and they are plotted
to show that they are strong enough to be well-fitted in all the blue boxes.
Bottom row: example where \ion{Mg}{7} 280.737\,\AA\, has the lowest intensity of all the boxes and the fit is not good.
}
  \label{fig7}
\end{figure*}

\subsection{Coronal rain density and temperature analysis}
\label{mgvii}
The \ion{Mg}{7} 280.737/278.402 ratio provides a density measurement at around 0.8\,MK, close to where the actual coronal rain is forming and is  
observable with EIS. A lower temperature constraint from \ion{Mg}{6} 270.394\,\AA, and surrounding temperature lines from \ion{Fe}{8} upwards,
allow us to attempt to isolate the DEM distribution for the coronal rain itself. Transition region emission is, however, weak in EIS observations
and this complicates the analysis. The \ion{Mg}{6} 270.394\,\AA\, line, for example, is blended on the long-wavelength side with 
the stronger \ion{Fe}{14} 270.519\,\AA\, line. Also, 
the \ion{Mg}{7} 280.737/278.402 ratio was discussed in 
application to transition region brightenings by \cite{Young2007} and they noted the presence of the \ion{Si}{7} 278.445\,\AA\, blend to 
\ion{Mg}{7} 278.402\,\AA.
Fortunately, both these blends are easily separable with a double Gaussian fit.
\cite{Young2007} also pointed out that the \ion{Mg}{7} 276.153\,\AA\, line could be
used to give an independent check on the deblending of \ion{Mg}{7} 278.402\,\AA\, since the \ion{Mg}{7} 276.153/278.402 branching ratio has a fixed
theoretical value. Unfortunately, this line is too weak in most of our boxes. 

Figure \ref{fig7} shows example line profiles for the weak Mg lines used in this analysis. The 3rd and 4th columns show the separation of the 
blended lines. In these columns, the Mg lines are the red curves and the blended lines are the light blue curves. The examples in this figure show where the limitations in the analysis are. As mentioned in Section \ref{overview}, the most
significant issue is the weak emission in \ion{Mg}{7} 280.737\,\AA. The figure contrasts a case where the \ion{Mg}{7} 280.737\,\AA\, emission in the
box is weak, but is still strong enough that the spectral fit is acceptable (top row), with a case where the emission is too weak to get a good fit
(bottom row). The figure also shows that the weakest Fe lines (\ion{Fe}{8} 185.213\,\AA\, and \ion{Fe}{9} 197.862\,\AA) are strong enough that
we can achieve a good fit in any box. 

\begin{figure*}
\centering
\includegraphics[viewport= 120 180 542 612,width=0.45\textwidth]{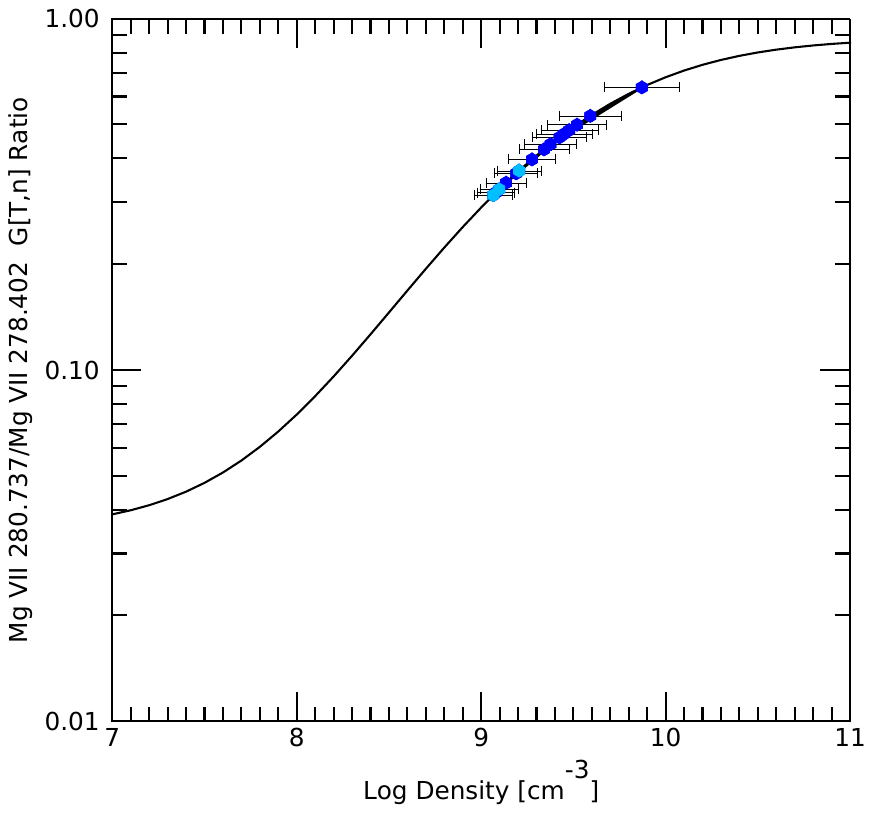}
\includegraphics[width=0.45\textwidth]{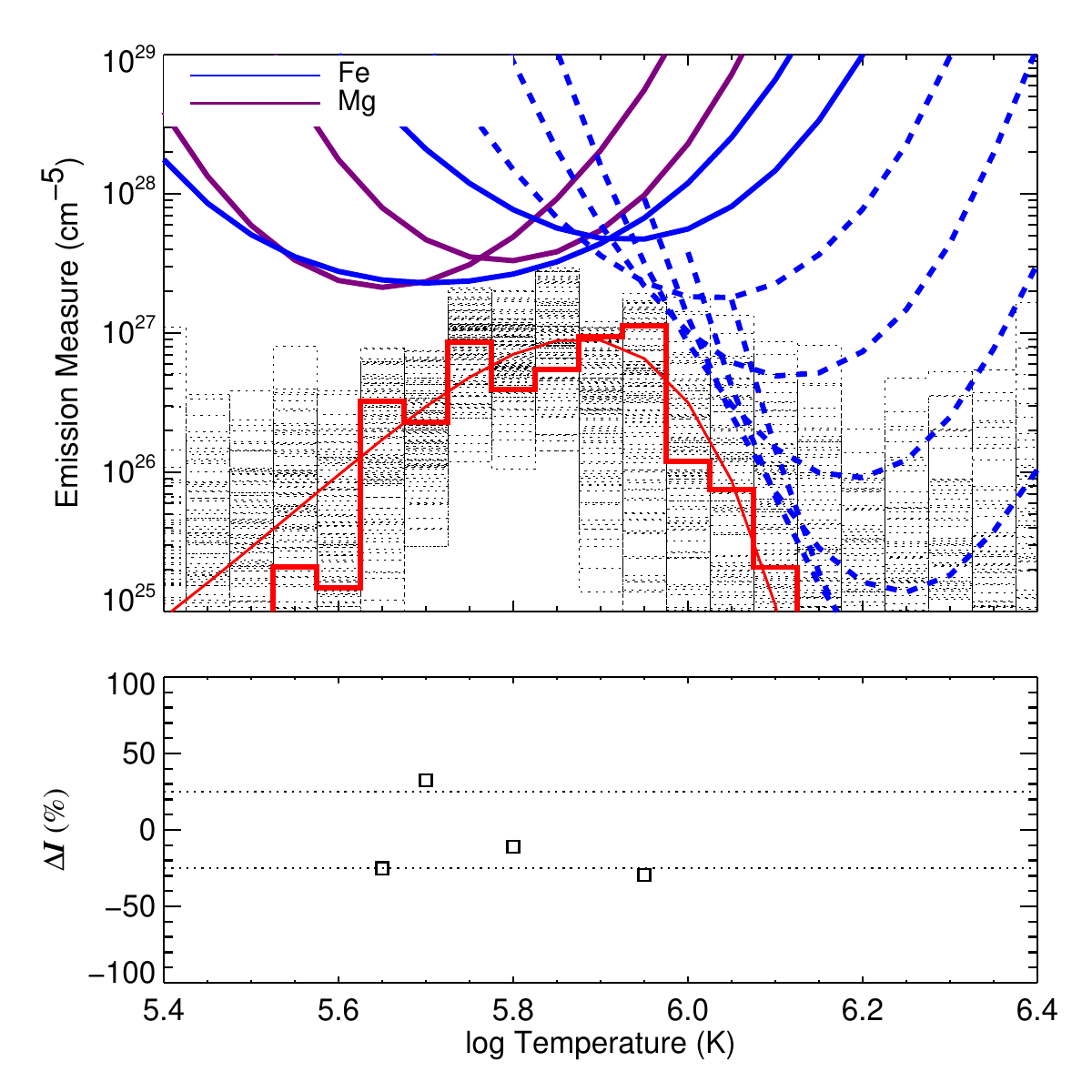}
\caption{ Coronal rain density and temperature analysis. Left panel: logarithmic electron density derived from the theoretical
\ion{Mg}{7} 280.737/278.402 line ratio. The ratio is shown by the solid black line. The blue dots indicate the measured
densities within 12/20 of the dark blue boxes shown in Figure \ref{fig2} and the light blue dots show the densities measured in the light blue boxes in Figure \ref{fig2}
that were added to the analysis to obtain extra \ion{Mg}{7} 280.737\,\AA\, measurements with a good S/N. 
The error bars show the range in densities that results from uncertainties in the 
observed ratio due to the radiometric calibration. 
Right panel (top): the emission measure (EM) distribution calculated
from the Mg and Fe lines where the intensity distributions are correlated with \ion{Mg}{7} 278.402\,\AA. 
The blue and magenta lines are EM loci curves that indicate the upper limits to the emission 
measure. The dashed lines are EM loci curves for lines where the intensity distributions are not correlated with \ion{Mg}{7} 278.402\,\AA.
The solid boxcar red line is the best fit solution from the MCMC algorithm. The dotted black lines show different 
realizations of the Monte Carlo solutions). 
The smooth thin red line is a Gaussian fit to the MCMC solution with a width
of $\sigma_T$ = 174,000\,K. 
Right panel (bottom): differences between the observed and calculated
intensities (expressed as a percentage) as a function of temperature. The dotted horizontal lines indicate an agreement
level of 25\%.
}
\label{fig8}
\end{figure*}

We found that an acceptable fit to the \ion{Mg}{7} 280.737\,\AA\, line was obtained in 12/20 dark blue boxes analyzed in the previous sections. To supplement
these measurements we added 3 more boxes to bring the total number to 15. The additional boxes are shown in light blue in Figure \ref{fig2}
and Figure \ref{fig8}. 
Figure \ref{fig8} (left panel) shows the theoretical \ion{Mg}{7} 280.737/278.402 intensity ratio plotted against density, the measurements in
the 12/20 boxes where the fits were acceptable as blue dots, and the 3 supplementary boxes as light blue dots. 
Densities in the coronal rain fall in the range of $\log (n/cm^{-3})$ = 9.1--9.9, with a mean value of $\log (n/cm^{-3})$ = 9.35.

The \ion{Mg}{7} 280.737\,\AA\, line is needed for the density measurement, but not for the DEM. For the DEM analysis, the weakest line is \ion{Mg}{6} 270.394\,\AA.
Acceptable fits were obtained for this line in 18/20 dark blue boxes. 
From Figure \ref{fig5} we can see that the EM distribution in the boxes is generally increasing with temperature, so the difficulty in isolating
the coronal rain component is deciding where the cut-off should be between the rain and the host loop. For coronal loops, we would generally
only include spectral lines that have highly correlated cross-field intensity profiles \citep[see e.g.][]{Aschwanden2008,Warren2008}. This is not
possible here, since the rain features are passing through the slit position and the boxes are too small for averaging along some spatial axis.
Instead, we looked at the correlation between spectral line intensities in an area encompassing all the boxes along the coronal rain time-distance trails. 
Lines were included if the linear Pearson correlation with the \ion{Mg}{7} 280.737\,\AA\, area intensity was higher than $r=0.6$. The intensities for the uncorrelated
lines were set to zero,
and the intensity error was set to 20\% of the observed intensity. This is a fairly
generous, low threshold that errs on the side of multi-thermality, however, the DEMs are narrow.

We show an example of the emission measure distribution from one of the boxes in the right panel of Figure \ref{fig8}. 
The EM peaks around 0.7\,MK. Only the \ion{Mg}{6} 270.394\,\AA, \ion{Mg}{7} 280.737\,\AA, \ion{Fe}{8} 185.213\,\AA, and \ion{Fe}{9} 197.862\.\AA\, lines 
meet the correlation threshold so the distribution is forced down rapidly at higher temperatures. In this example, all the observed intensities
are reproduced to within 30\%. The best-fit MCMC solution (thick red line) also shows that
the EM distribution is narrow. We have overplotted a Gaussian fit with a width of $\sigma_T$ = 174,000\,K to the MCMC solution. This is narrower
than most warm coronal loops \citep{Warren2008}, and the MCMC solution is narrower still.

\subsection{Potential sources of non-thermal line broadening}  
\label{broadening}
In Section \ref{velocities} we found an increase in non-thermal velocities with decreasing temperatures. This result could indicate 
the presence of turbulence, or it could result from pressure broadening or opacity broadening.

Following \cite{Milligan2011}, we can estimate the amount of pressure broadening from 
\begin{equation}
\delta \lambda \sim {\lambda^2  \over c} {n \sigma \over \pi} \sqrt {( 2 k_B T/M )}
\label{eq2}
\end{equation} 
where $\lambda$ is the wavelength of the spectral line, $c$ is the speed of light, $n$ is the electron density, $\sigma$ is the collisional ionisation
cross-section, $T$ is the formation temperature, $k_B$ is Boltzmann's constant, and $M$ is the ion mass. \cite{Milligan2011} showed that $\delta \lambda \sim$
10$^{-15}$\,\AA\, for the \ion{Fe}{14} 264.787\,\AA\, and \ion{Fe}{14} 274.203\,\AA\, lines used in his study of the hard X-ray footpoints of a solar
flare and was therefore negligible. \cite{Milligan2011} assumed a density of 10$^{11}$ cm$^{-3}$ for this calculation. 
Since $\delta \lambda$ depends linearly on density in Equation \ref{eq2}, we expect that the contribution from pressure
broadening is even smaller for these lines in our host loops (mean $\log (n/cm^{-3})$=8.65 from the \ion{Fe}{13} ratio - see Section \ref{dem}).
At the lowest temperatures where the non-thermal velocities were measured using \ion{Fe}{8} 185.213\,\AA, the densities are higher (mean $\log (n/cm^{-3})$=9.35 from the \ion{Mg}{7} ratio - see Section \ref{mgvii}),
but are still over an order of magnitude lower than 10$^{11}$ cm$^{-3}$. Some other factors in Equation \ref{eq2} ($\lambda$, $T$, $\sigma$) do change, but
their effect is to cancel each other: $\lambda$ decreases from 264.787\,\AA\, to 185.213\,\AA, $T$ decreases from 2\,MK to 0.45\,MK, and $\sigma$ increases from e.g. 5 to 35 $\times$10$^{-19}$\,cm$^{-2}$ at 1000\,eV \citep{Hahn2011,Hahn2015}. We conclude that pressure broadening does not impact our measurements.

The optical depth of a spectral line at line center can be expressed as given by \cite{Mitchell1961},
\begin{equation}
\tau_0 = 1.16 \times 10^{-6} \sqrt {M/T} \lambda N_l f_{lu} h
\end{equation}
where $f_{lu}$ is the absorption oscillator strength for the transition from the lower ($l$) to the upper ($u$) level, $h$ is the path length, and 
$N_l$ is the population of the lower level. Using oscillator strengths from the CHIANTI v.10 database, and adopting the coronal element abundances
of \cite{Feldman1992} together with the CHIANTI ionisation fractions (to calculate $N_l$), we computed values of $\tau_0$ for \ion{Fe}{13} 202.044\,\AA\,
and \ion{Fe}{8} 186.601\,\AA\, using the densities quoted above for the closest temperature. Both of these lines arise from transitions that 
terminate in the ground state of their respective ions (so $N_l$ = the ground state).
We assume a typical value of 300\,km for the size of the coronal rain clump.
We find that for \ion{Fe}{13} 202.044\,\AA, $\tau_0$ = 0.003, and 
for \ion{Fe}{8} 186.601\,\AA, $\tau_0$ = 0.2, indicating that both lines are optically thin and therefore opacity broadening is not significant.

Though it is unclear if we should expect the rain clumps to be more turbulent that the
surrounding host loops, with pressure and opacity broadening ruled out, and lacking any other obvious explanations, we suggest that the broadening is a result of 
turbulence as the rain condensations form.
\section{Discussion and Conclusions}
We have applied several spectroscopic diagnostics to a rare set of EIS sit-and-stare observations of post-flare loop
cooling and the formation and evolution of coronal rain showers, including when the rain is first condensing. 
The main observational results suggest the following picture.
Quiescent coronal rain is present in the active region before the X2.1 flare. The flare, however, destabilizes and energises
the loop arcade, and EIS observes while the loop system is relaxing to its pre-flare state. We observe a coronal composition
in the high temperature ($\sim$ 3.5\,MK) loops, that then cool down to $\sim$ 1.3\,MK, where we measure typical densities
of $\log (n/cm^{-3}) \sim$ 8.65. At these temperatures, the plasma composition is photospheric. As the loops cool further,
coronal rain forms, with higher densities of $\log (n/cm^{-3}) \sim$ 9.35 reached at 0.7\,MK. Non-thermal velocities also
increase as the temperature decreases. The temperature distributions
in the rain showers are narrow. Coronal rain is visible throughout the observations, and is seen in loops disrupted by the flare close
to the impulsive phase.
These are important constraints for numerical modeling and understanding the formation of coronal rain.

The bifurcation in plasma composition measurements at the two temperatures does suggest we are observing post-flare driven coronal rain.
In quiescent conditions, active region loops should show a coronal composition in both the Si/S and Ca/Ar diagnostic. 
Based on the ponderomotive force model of the FIP effect \citep{Laming2004,Laming2015}, where the forces are induced by reflection and refraction of Alfv\'{e}n waves
generated by coronal (nanoflare) reconnection, low FIP (ionized) material is preferentially accelerated from the top of the chromosphere \citep{Brooks2021b}, and this qualitatively reproduces
the expected behavior of Si/S and Ca/Ar in AR loops. 
It is unclear by what process a coronal composition loop could develop a photospheric composition as it cools.
In flaring conditions,
however, the energy is sufficient to ablate plasma from the deep chromosphere. There S acts like a low-FIP element, so that it fractionates in a similar
manner to Si, and no FIP effect is detected in the Si/S ratio. The FIP effect is, however, still detected in Ca/Ar, and this naturally
explains what we see here. This difference in Si/S and Ca/Ar is in effect a diagnostic of the source of the coronal plasma because S behaves differently depending on
the fractionation location \citep{Laming2019}.

The increase in non-thermal velocities and densities with decreasing temperature suggests the development of turbulence as the coronal rain
condensations form in the host loops. Our direct measurements of densities in the rain showers are somewhat lower than previously reported
in the chromosphere.
This is likely due to the uncertainties in inferring densities indirectly from imaging data, and also because we are making measurements
early in the formation at higher temperatures. The densities are likely to grow as the instability and thermal runaway continues.

The transition to loop emission at the temperature of \ion{Fe}{10} suggests rain should start to form $\sim$ 1.1\,MK.
The formation of coronal rain is also influenced by the magnitude of the radiative losses, since they increase as the plasma cools, 
and these are affected by the plasma composition. In quiescent rain, the host loops have a coronal composition and typical
radiative loss function with S behaving like a high-FIP element (S would behave this way even if the loops had a photospheric composition).
In the flare-driven rain we observe here, however, once the host loops have cooled to around 1\,MK,
the radiative losses are increased. This is because, as discussed above, S behaves like a low-FIP element and is enhanced 
in addition to all the other low FIP elements. The increased radiative losses thus
make it easier for rain to form and cause it to cool more rapidly. This difference in radiative losses, driven by the behavior of S, 
is an essential difference between flare-driven and quiescent rain. 

Coronal rain appears to show coherent behavior in imaging data, leading to the use of the term rain
showers. The narrow temperature distributions measured by EIS in the rain showers support this idea of rain features evolving coherently.
This is also reminiscent of the
collective evolution of strands in the overlying coronal loops themselves \citep{Warren2008}, suggesting that rain showers might correspond
to the host loops, and rain blobs might correspond to the loop threads. 
There is some supporting evidence for this scenario from numerical simulations, which show that rain forms initially on one strand, but clearly
occurs across multiple strands \citep{Antolin2022b,Li2022}.

The spatio-temporal resolution of EIS is not sufficient to 
resolve either the 200-300\,km average size of the rain blobs, or loop strands. Modeling of the cross-loop intensity profiles, however, suggests that a future instrument with a spatial
resolution of 200\,km\, should be able to resolve coronal loops \citep{Brooks2012,Brooks2013}. This spatial resolution will be achieved by
the Solar-C Extreme UltraViolet high-throughput Spectroscopic Telescope \citep[EUVST,][]{Shimizu2020}. 
Our study shows the value of spectroscopic observations for gaining insights into properties of the heating mechanism
from the loop cooling phase. The spatial resolution of Solar-C EUVST
will be sufficient to resolve coronal loop threads and rain blobs, and the seamless broad temperature coverage will allow tracing of their evolution from the corona to the chromosphere.

\begin{acknowledgments}
We thank the referee for the thorough review of the article and many suggestions for improvement.
D.H.B. thanks Dr Patrick Antolin for helpful discussions and comments.
This work was funded by the NASA Hinode program. 
J.W.R., I.U.U., and H.P.W. were also supported by a NASA Heliophysics Supporting Research Grant, number NNH19ZDA001N.
Hinode is a Japanese mission developed and launched by ISAS/JAXA, collaborating with NAOJ as a domestic partner, NASA and STFC (UK) as international partners. Scientific operation of the Hinode mission is conducted by the Hinode science team organized at ISAS/JAXA. This team mainly consists of scientists from institutes in the partner countries. Support for the post-launch operation is provided by JAXA and NAOJ(Japan), STFC (U.K.), NASA, ESA, and NSC (Norway).
The AIA data are courtesy of NASA/SDO and the AIA, EVE, and HMI science teams.
CHIANTI is a collaborative project involving George Mason University, the University of Michigan (USA), University of Cambridge (UK) and NASA Goddard Space Flight Center (USA).

\end{acknowledgments}

\facilities{SDO/AIA, \emph{Hinode}/EIS}


\begin{thebibliography}{}
\expandafter\ifx\csname natexlab\endcsname\relax\def\natexlab#1{#1}\fi
\providecommand{\url}[1]{\href{#1}{#1}}

\bibitem[{{Ahn} {et~al.}(2014){Ahn}, {Chae}, {Cho}, {Song}, {Yang}, {Goode},
  {Cao}, {Park}, {Nah}, {Jang}, \& {Park}}]{Ahn2014}
{Ahn}, K., {Chae}, J., {Cho}, K.-S., {et~al.} 2014, \solphys, 289, 4117

\bibitem[{{Antiochos} {et~al.}(1999){Antiochos}, {MacNeice}, {Spicer}, \&
  {Klimchuk}}]{Antiochos1999}
{Antiochos}, S.~K., {MacNeice}, P.~J., {Spicer}, D.~S., \& {Klimchuk}, J.~A.
  1999, \apj, 512, 985

\bibitem[{{Antolin}(2020)}]{Antolin2020}
{Antolin}, P. 2020, Plasma Physics and Controlled Fusion, 62, 014016

\bibitem[{{Antolin} \& {Froment}(2022)}]{Antolin2022a}
{Antolin}, P., \& {Froment}, C. 2022, Frontiers in Astronomy and Space
  Sciences, 9, 820116

\bibitem[{{Antolin} {et~al.}(2022){Antolin}, {Mart{\'\i}nez-Sykora}, \&
  {{\c{S}}ahin}}]{Antolin2022b}
{Antolin}, P., {Mart{\'\i}nez-Sykora}, J., \& {{\c{S}}ahin}, S. 2022, \apjl,
  926, L29

\bibitem[{{Antolin} \& {Rouppe van der Voort}(2012)}]{Antolin2012}
{Antolin}, P., \& {Rouppe van der Voort}, L. 2012, \apj, 745, 152

\bibitem[{{Antolin} {et~al.}(2010){Antolin}, {Shibata}, \&
  {Vissers}}]{Antolin2010}
{Antolin}, P., {Shibata}, K., \& {Vissers}, G. 2010, \apj, 716, 154

\bibitem[{{Antolin} {et~al.}(2015){Antolin}, {Vissers}, {Pereira}, {Rouppe van
  der Voort}, \& {Scullion}}]{Antolin2015}
{Antolin}, P., {Vissers}, G., {Pereira}, T.~M.~D., {Rouppe van der Voort}, L.,
  \& {Scullion}, E. 2015, \apj, 806, 81

\bibitem[{{Aschwanden} {et~al.}(2008){Aschwanden}, {Nitta}, {Wuelser}, \&
  {Lemen}}]{Aschwanden2008}
{Aschwanden}, M.~J., {Nitta}, N.~V., {Wuelser}, J.-P., \& {Lemen}, J.~R. 2008,
  \apj, 680, 1477

\bibitem[{{Baker} {et~al.}(2019){Baker}, {van Driel-Gesztelyi}, {Brooks},
  {Valori}, {James}, {Laming}, {Long}, {D{\'e}moulin}, {Green}, {Matthews},
  {Ol{\'a}h}, \& {K{\H{o}}v{\'a}ri}}]{Baker2019}
{Baker}, D., {van Driel-Gesztelyi}, L., {Brooks}, D.~H., {et~al.} 2019, \apj,
  875, 35

\bibitem[{{Brooks} {et~al.}(2015){Brooks}, {Ugarte-Urra}, \&
  {Warren}}]{Brooks2015}
{Brooks}, D.~H., {Ugarte-Urra}, I., \& {Warren}, H.~P. 2015, Nature
  Communications, 6, 5947

\bibitem[{{Brooks} \& {Warren}(2011)}]{Brooks2011}
{Brooks}, D.~H., \& {Warren}, H.~P. 2011, \apjl, 727, L13

\bibitem[{{Brooks} \& {Warren}(2012)}]{Brooks2012}
---. 2012, \apjl, 760, L5

\bibitem[{{Brooks} \& {Warren}(2016)}]{Brooks2016}
---. 2016, \apj, 820, 63

\bibitem[{{Brooks} {et~al.}(2013){Brooks}, {Warren}, {Ugarte-Urra}, \&
  {Winebarger}}]{Brooks2013}
{Brooks}, D.~H., {Warren}, H.~P., {Ugarte-Urra}, I., \& {Winebarger}, A.~R.
  2013, \apjl, 772, L19

\bibitem[{{Brooks} \& {Yardley}(2021)}]{Brooks2021b}
{Brooks}, D.~H., \& {Yardley}, S.~L. 2021, Science Advances, 7, eabf0068

\bibitem[{{Brooks} {et~al.}(2020){Brooks}, {Winebarger}, {Savage}, {Warren},
  {De Pontieu}, {Peter}, {Cirtain}, {Golub}, {Kobayashi}, {McIntosh},
  {McKenzie}, {Morton}, {Rachmeler}, {Testa}, {Tiwari}, \&
  {Walsh}}]{Brooks2020}
{Brooks}, D.~H., {Winebarger}, A.~R., {Savage}, S., {et~al.} 2020, \apj, 894,
  144

\bibitem[{{{\c{S}}ahin} \& {Antolin}(2022)}]{Sahin2022}
{{\c{S}}ahin}, S., \& {Antolin}, P. 2022, \apjl, 931, L27

\bibitem[{{Culhane} {et~al.}(2007){Culhane}, {Harra}, {James}, {Al-Janabi},
  {Bradley}, {Chaudry}, {Rees}, {Tandy}, {Thomas}, {Whillock}, {Winter},
  {Doschek}, {Korendyke}, {Brown}, {Myers}, {Mariska}, {Seely}, {Lang}, {Kent},
  {Shaughnessy}, {Young}, {Simnett}, {Castelli}, {Mahmoud}, {Mapson-Menard},
  {Probyn}, {Thomas}, {Davila}, {Dere}, {Windt}, {Shea}, {Hagood}, {Moye},
  {Hara}, {Watanabe}, {Matsuzaki}, {Kosugi}, {Hansteen}, \&
  {Wikstol}}]{Culhane2007}
{Culhane}, J.~L., {Harra}, L.~K., {James}, A.~M., {et~al.} 2007, Sol. Phys.,
  243, 19

\bibitem[{{De Pontieu} {et~al.}(2009){De Pontieu}, {McIntosh}, {Hansteen}, \&
  {Schrijver}}]{DePontieu2009}
{De Pontieu}, B., {McIntosh}, S.~W., {Hansteen}, V.~H., \& {Schrijver}, C.~J.
  2009, \apjl, 701, L1

\bibitem[{{De Pontieu} {et~al.}(2011){De Pontieu}, {McIntosh}, {Carlsson},
  {Hansteen}, {Tarbell}, {Boerner}, {Martinez-Sykora}, {Schrijver}, \&
  {Title}}]{DePontieu2011}
{De Pontieu}, B., {McIntosh}, S.~W., {Carlsson}, M., {et~al.} 2011, Science,
  331, 55

\bibitem[{{Del Zanna} {et~al.}(2021){Del Zanna}, {Dere}, {Young}, \&
  {Landi}}]{DelZanna2021}
{Del Zanna}, G., {Dere}, K.~P., {Young}, P.~R., \& {Landi}, E. 2021, \apj, 909,
  38

\bibitem[{{Del Zanna} {et~al.}(2023){Del Zanna}, {Weberg}, \&
  {Warren}}]{DelZanna2023}
{Del Zanna}, G., {Weberg}, M., \& {Warren}, H.~P. 2023, arXiv e-prints,
  arXiv:2308.06609

\bibitem[{{Dere} {et~al.}(1997){Dere}, {Landi}, {Mason}, {Monsignori Fossi}, \&
  {Young}}]{Dere1997}
{Dere}, K.~P., {Landi}, E., {Mason}, H.~E., {Monsignori Fossi}, B.~C., \&
  {Young}, P.~R. 1997, \aaps, 125, 149

\bibitem[{{Doschek} {et~al.}(2015){Doschek}, {Warren}, \&
  {Feldman}}]{Doschek2015}
{Doschek}, G.~A., {Warren}, H.~P., \& {Feldman}, U. 2015, \apjl, 808, L7

\bibitem[{{Downs} {et~al.}(2016){Downs}, {Lionello}, {Miki{\'c}}, {Linker}, \&
  {Velli}}]{Downs2016}
{Downs}, C., {Lionello}, R., {Miki{\'c}}, Z., {Linker}, J.~A., \& {Velli}, M.
  2016, \apj, 832, 180

\bibitem[{{Feldman}(1992)}]{Feldman1992}
{Feldman}, U. 1992, \physscr, 46, 202

\bibitem[{{Freeland} \& {Handy}(1998)}]{Freeland1998}
{Freeland}, S.~L., \& {Handy}, B.~N. 1998, \solphys, 182, 497

\bibitem[{{Froment} {et~al.}(2020){Froment}, {Antolin}, {Henriques},
  {Kohutova}, \& {Rouppe van der Voort}}]{Froment2020}
{Froment}, C., {Antolin}, P., {Henriques}, V.~M.~J., {Kohutova}, P., \& {Rouppe
  van der Voort}, L.~H.~M. 2020, \aap, 633, A11

\bibitem[{{Froment} {et~al.}(2015){Froment}, {Auch{\`e}re}, {Bocchialini},
  {Buchlin}, {Guennou}, \& {Solomon}}]{Froment2015}
{Froment}, C., {Auch{\`e}re}, F., {Bocchialini}, K., {et~al.} 2015, \apj, 807,
  158

\bibitem[{{Froment} {et~al.}(2018){Froment}, {Auch{\`e}re}, {Miki{\'c}},
  {Aulanier}, {Bocchialini}, {Buchlin}, {Solomon}, \&
  {Soubri{\'e}}}]{Froment2018}
{Froment}, C., {Auch{\`e}re}, F., {Miki{\'c}}, Z., {et~al.} 2018, \apj, 855, 52

\bibitem[{{Gold} \& {Hoyle}(1960)}]{Gold1960}
{Gold}, T., \& {Hoyle}, F. 1960, \mnras, 120, 89

\bibitem[{{Hahn} {et~al.}(2011){Hahn}, {Grieser}, {Krantz}, {Lestinsky},
  {M{\"u}ller}, {Novotn{\'y}}, {Repnow}, {Schippers}, {Wolf}, \&
  {Savin}}]{Hahn2011}
{Hahn}, M., {Grieser}, M., {Krantz}, C., {et~al.} 2011, \apj, 735, 105

\bibitem[{{Hahn} {et~al.}(2015){Hahn}, {Becker}, {Bernhardt}, {Grieser},
  {Krantz}, {Lestinsky}, {M{\"u}ller}, {Novotn{\'y}}, {Repnow}, {Schippers},
  {Spruck}, {Wolf}, \& {Savin}}]{Hahn2015}
{Hahn}, M., {Becker}, A., {Bernhardt}, D., {et~al.} 2015, \apj, 813, 16

\bibitem[{{Jing} {et~al.}(2016){Jing}, {Xu}, {Cao}, {Liu}, {Gary}, \&
  {Wang}}]{Jing2016}
{Jing}, J., {Xu}, Y., {Cao}, W., {et~al.} 2016, Scientific Reports, 6, 24319

\bibitem[{{Kamio} {et~al.}(2010){Kamio}, {Hara}, {Watanabe}, {Fredvik}, \&
  {Hansteen}}]{Kamio2010}
{Kamio}, S., {Hara}, H., {Watanabe}, T., {Fredvik}, T., \& {Hansteen}, V.~H.
  2010, \solphys, 266, 209

\bibitem[{{Karpen} {et~al.}(2001){Karpen}, {Antiochos}, {Hohensee}, {Klimchuk},
  \& {MacNeice}}]{Karpen2001}
{Karpen}, J.~T., {Antiochos}, S.~K., {Hohensee}, M., {Klimchuk}, J.~A., \&
  {MacNeice}, P.~J. 2001, \apjl, 553, L85

\bibitem[{{Kashyap} \& {Drake}(1998)}]{Kashyap1998}
{Kashyap}, V., \& {Drake}, J.~J. 1998, \apj, 503, 450

\bibitem[{{Kashyap} \& {Drake}(2000)}]{Kashyap2000}
---. 2000, Bulletin of the Astronomical Society of India, 28, 475

\bibitem[{{Kleint} {et~al.}(2014){Kleint}, {Antolin}, {Tian}, {Judge}, {Testa},
  {De Pontieu}, {Mart{\'\i}nez-Sykora}, {Reeves}, {Wuelser}, {McKillop},
  {Saar}, {Carlsson}, {Boerner}, {Hurlburt}, {Lemen}, {Tarbell}, {Title},
  {Golub}, {Hansteen}, {Jaeggli}, \& {Kankelborg}}]{Kleint2014}
{Kleint}, L., {Antolin}, P., {Tian}, H., {et~al.} 2014, \apjl, 789, L42

\bibitem[{{Klimchuk}(2006)}]{Klimchuk2006}
{Klimchuk}, J.~A. 2006, \solphys, 234, 41

\bibitem[{{Kohutova} {et~al.}(2020){Kohutova}, {Antolin}, {Popovas},
  {Szydlarski}, \& {Hansteen}}]{Kohutova2020}
{Kohutova}, P., {Antolin}, P., {Popovas}, A., {Szydlarski}, M., \& {Hansteen},
  V.~H. 2020, \aap, 639, A20

\bibitem[{{Kohutova} \& {Verwichte}(2016)}]{Kohutova2016}
{Kohutova}, P., \& {Verwichte}, E. 2016, \apj, 827, 39

\bibitem[{{Kohutova} {et~al.}(2019){Kohutova}, {Verwichte}, \&
  {Froment}}]{Kohutova2019}
{Kohutova}, P., {Verwichte}, E., \& {Froment}, C. 2019, \aap, 630, A123

\bibitem[{{Kosugi} {et~al.}(2007){Kosugi}, {Matsuzaki}, {Sakao}, {Shimizu},
  {Sone}, {Tachikawa}, {Hashimoto}, {Minesugi}, {Ohnishi}, {Yamada}, {Tsuneta},
  {Hara}, {Ichimoto}, {Suematsu}, {Shimojo}, {Watanabe}, {Shimada}, {Davis},
  {Hill}, {Owens}, {Title}, {Culhane}, {Harra}, {Doschek}, \&
  {Golub}}]{Kosugi2007}
{Kosugi}, T., {Matsuzaki}, K., {Sakao}, T., {et~al.} 2007, Sol. Phys., 243, 3

\bibitem[{{Laming}(2004)}]{Laming2004}
{Laming}, J.~M. 2004, \apj, 614, 1063

\bibitem[{{Laming}(2015)}]{Laming2015}
---. 2015, Living Reviews in Solar Physics, 12, 2

\bibitem[{{Laming} {et~al.}(2019){Laming}, {Vourlidas}, {Korendyke}, {Chua},
  {Cranmer}, {Ko}, {Kuroda}, {Provornikova}, {Raymond}, {Raouafi}, {Strachan},
  {Tun-Beltran}, {Weberg}, \& {Wood}}]{Laming2019}
{Laming}, J.~M., {Vourlidas}, A., {Korendyke}, C., {et~al.} 2019, \apj, 879,
  124

\bibitem[{{Lang} {et~al.}(2006){Lang}, {Kent}, {Paustian}, {Brown}, {Keyser},
  {Anderson}, {Case}, {Chaudry}, {James}, {Korendyke}, {Pike}, {Probyn},
  {Rippington}, {Seely}, {Tandy}, \& {Whillock}}]{lang2006}
{Lang}, J., {Kent}, B.~J., {Paustian}, W., {et~al.} 2006, \ao, 45, 8689

\bibitem[{{Lemen} {et~al.}(2012){Lemen}, {Title}, {Akin}, {Boerner}, {Chou},
  {Drake}, {Duncan}, {Edwards}, {Friedlaender}, {Heyman}, {Hurlburt}, {Katz},
  {Kushner}, {Levay}, {Lindgren}, {Mathur}, {McFeaters}, {Mitchell}, {Rehse},
  {Schrijver}, {Springer}, {Stern}, {Tarbell}, {Wuelser}, {Wolfson}, {Yanari},
  {Bookbinder}, {Cheimets}, {Caldwell}, {Deluca}, {Gates}, {Golub}, {Park},
  {Podgorski}, {Bush}, {Scherrer}, {Gummin}, {Smith}, {Auker}, {Jerram},
  {Pool}, {Soufli}, {Windt}, {Beardsley}, {Clapp}, {Lang}, \&
  {Waltham}}]{Lemen2012}
{Lemen}, J.~R., {Title}, A.~M., {Akin}, D.~J., {et~al.} 2012, Sol. Phys., 275,
  17

\bibitem[{{Li} {et~al.}(2022){Li}, {Keppens}, \& {Zhou}}]{Li2022}
{Li}, X., {Keppens}, R., \& {Zhou}, Y. 2022, \apj, 926, 216

\bibitem[{{Milligan}(2011)}]{Milligan2011}
{Milligan}, R.~O. 2011, \apj, 740, 70

\bibitem[{{Mitchell} \& {Zemansky}(1961)}]{Mitchell1961}
{Mitchell}, A.~C.~G., \& {Zemansky}, M.~W. 1961, {Resonance Radiation and
  Excited Atoms, Cambridge University Press}

\bibitem[{{M{\"u}ller} {et~al.}(2005){M{\"u}ller}, {De Groof}, {Hansteen}, \&
  {Peter}}]{Muller2005}
{M{\"u}ller}, D.~A.~N., {De Groof}, A., {Hansteen}, V.~H., \& {Peter}, H. 2005,
  \aap, 436, 1067

\bibitem[{{M{\"u}ller} {et~al.}(2003){M{\"u}ller}, {Hansteen}, \&
  {Peter}}]{Muller2003}
{M{\"u}ller}, D.~A.~N., {Hansteen}, V.~H., \& {Peter}, H. 2003, \aap, 411, 605

\bibitem[{{Oliver} {et~al.}(2014){Oliver}, {Soler}, {Terradas}, {Zaqarashvili},
  \& {Khodachenko}}]{Oliver2014}
{Oliver}, R., {Soler}, R., {Terradas}, J., {Zaqarashvili}, T.~V., \&
  {Khodachenko}, M.~L. 2014, \apj, 784, 21

\bibitem[{{Parker}(1983)}]{Parker1983}
{Parker}, E.~N. 1983, \apj, 264, 642

\bibitem[{{Parker}(1988)}]{Parker1988}
---. 1988, \apj, 330, 474

\bibitem[{{Parnell} \& {De Moortel}(2012)}]{Parnell2012}
{Parnell}, C.~E., \& {De Moortel}, I. 2012, Philosophical Transactions of the
  Royal Society of London Series A, 370, 3217

\bibitem[{{Pesnell} {et~al.}(2012){Pesnell}, {Thompson}, \&
  {Chamberlin}}]{Pesnell2012}
{Pesnell}, W.~D., {Thompson}, B.~J., \& {Chamberlin}, P.~C. 2012, Sol. Phys.,
  275, 3

\bibitem[{{Peter}(2001)}]{Peter2001}
{Peter}, H. 2001, \aap, 374, 1108

\bibitem[{{Pontin} \& {Priest}(2022)}]{Pontin2022}
{Pontin}, D.~I., \& {Priest}, E.~R. 2022, Living Reviews in Solar Physics, 19,
  1

\bibitem[{{Reale}(2014)}]{Reale2014}
{Reale}, F. 2014, Living Reviews in Solar Physics, 11, 4

\bibitem[{{Reep} {et~al.}(2020){Reep}, {Antolin}, \& {Bradshaw}}]{Reep2020}
{Reep}, J.~W., {Antolin}, P., \& {Bradshaw}, S.~J. 2020, \apj, 890, 100

\bibitem[{{Scharmer} {et~al.}(2008){Scharmer}, {Narayan}, {Hillberg}, {de la
  Cruz Rodriguez}, {L{\"o}fdahl}, {Kiselman}, {S{\"u}tterlin}, {van Noort}, \&
  {Lagg}}]{Scharmer2008}
{Scharmer}, G.~B., {Narayan}, G., {Hillberg}, T., {et~al.} 2008, \apjl, 689,
  L69

\bibitem[{{Schrijver}(2001)}]{Schrijver2001}
{Schrijver}, C.~J. 2001, \solphys, 198, 325

\bibitem[{{Scott} {et~al.}(2015{\natexlab{a}}){Scott}, {Asplund}, {Grevesse},
  {Bergemann}, \& {Sauval}}]{Scott2015b}
{Scott}, P., {Asplund}, M., {Grevesse}, N., {Bergemann}, M., \& {Sauval}, A.~J.
  2015{\natexlab{a}}, \aap, 573, A26

\bibitem[{{Scott} {et~al.}(2015{\natexlab{b}}){Scott}, {Grevesse}, {Asplund},
  {Sauval}, {Lind}, {Takeda}, {Collet}, {Trampedach}, \& {Hayek}}]{Scott2015a}
{Scott}, P., {Grevesse}, N., {Asplund}, M., {et~al.} 2015{\natexlab{b}}, \aap,
  573, A25

\bibitem[{{Scullion} {et~al.}(2016){Scullion}, {Rouppe van der Voort},
  {Antolin}, {Wedemeyer}, {Vissers}, {Kontar}, \& {Gallagher}}]{Scullion2016}
{Scullion}, E., {Rouppe van der Voort}, L., {Antolin}, P., {et~al.} 2016, \apj,
  833, 184

\bibitem[{{Shay} {et~al.}(2001){Shay}, {Drake}, {Rogers}, \&
  {Denton}}]{Shay2001}
{Shay}, M.~A., {Drake}, J.~F., {Rogers}, B.~N., \& {Denton}, R.~E. 2001, \jgr,
  106, 3759

\bibitem[{{Shimizu} {et~al.}(2020){Shimizu}, {Imada}, {Kawate}, {Suematsu},
  {Hara}, {Tsuzuki}, {Katsukawa}, {Kubo}, {Ishikawa}, {Watanabe}, {Toriumi},
  {Ichimoto}, {Nagata}, {Hasegawa}, {Yokoyama}, {Watanabe}, {Tsuno},
  {Korendyke}, {Warren}, {De Pontieu}, {Boerner}, {Solanki}, {Teriaca},
  {Schuehle}, {Matthews}, {Long}, {Thomas}, {Hancock}, {Reid}, {Fludra},
  {Auch{\`e}re}, {Andretta}, {Naletto}, {Poletto}, \& {Harra}}]{Shimizu2020}
{Shimizu}, T., {Imada}, S., {Kawate}, T., {et~al.} 2020, in Society of
  Photo-Optical Instrumentation Engineers (SPIE) Conference Series, Vol. 11444,
  Space Telescopes and Instrumentation 2020: Ultraviolet to Gamma Ray, ed.
  J.-W.~A. {den Herder}, S.~{Nikzad}, \& K.~{Nakazawa}, 114440N

\bibitem[{{Ugarte-Urra} {et~al.}(2009){Ugarte-Urra}, {Warren}, \&
  {Brooks}}]{Ugarte2009}
{Ugarte-Urra}, I., {Warren}, H.~P., \& {Brooks}, D.~H. 2009, \apj, 695, 642

\bibitem[{{van Ballegooijen} {et~al.}(2011){van Ballegooijen}, {Asgari-Targhi},
  {Cranmer}, \& {DeLuca}}]{vanBallegooijen2011}
{van Ballegooijen}, A.~A., {Asgari-Targhi}, M., {Cranmer}, S.~R., \& {DeLuca},
  E.~E. 2011, \apj, 736, 3

\bibitem[{{Warren}(2014)}]{Warren2014a}
{Warren}, H.~P. 2014, \apjl, 786, L2

\bibitem[{{Warren} {et~al.}(2016){Warren}, {Brooks}, {Doschek}, \&
  {Feldman}}]{Warren2016}
{Warren}, H.~P., {Brooks}, D.~H., {Doschek}, G.~A., \& {Feldman}, U. 2016,
  \apj, 824, 56

\bibitem[{{Warren} {et~al.}(2008){Warren}, {Ugarte-Urra}, {Doschek}, {Brooks},
  \& {Williams}}]{Warren2008}
{Warren}, H.~P., {Ugarte-Urra}, I., {Doschek}, G.~A., {Brooks}, D.~H., \&
  {Williams}, D.~R. 2008, \apjl, 686, L131

\bibitem[{{Warren} {et~al.}(2014){Warren}, {Ugarte-Urra}, \&
  {Landi}}]{Warren2014}
{Warren}, H.~P., {Ugarte-Urra}, I., \& {Landi}, E. 2014, \apjs, 213, 11

\bibitem[{{Warren} {et~al.}(2011){Warren}, {Ugarte-Urra}, {Young}, \&
  {Stenborg}}]{Warren2011}
{Warren}, H.~P., {Ugarte-Urra}, I., {Young}, P.~R., \& {Stenborg}, G. 2011,
  \apj, 727, 58

\bibitem[{{Watanabe} {et~al.}(2012){Watanabe}, {Masuda}, \&
  {Segawa}}]{Watanabe2012}
{Watanabe}, K., {Masuda}, S., \& {Segawa}, T. 2012, \solphys, 279, 317

\bibitem[{{Wedemeyer-B{\"o}hm} {et~al.}(2012){Wedemeyer-B{\"o}hm}, {Scullion},
  {Steiner}, {Rouppe van der Voort}, {de La Cruz Rodriguez}, {Fedun}, \&
  {Erd{\'e}lyi}}]{Wedemeyer2012}
{Wedemeyer-B{\"o}hm}, S., {Scullion}, E., {Steiner}, O., {et~al.} 2012, \nat,
  486, 505

\bibitem[{{Winebarger} \& {Warren}(2004)}]{Winebarger2004}
{Winebarger}, A.~R., \& {Warren}, H.~P. 2004, \apjl, 610, L129

\bibitem[{{Young} {et~al.}(2007){Young}, {Del Zanna}, {Mason}, {Doschek},
  {Culhane}, \& {Hara}}]{Young2007}
{Young}, P.~R., {Del Zanna}, G., {Mason}, H.~E., {et~al.} 2007, \pasj, 59, S727

\end{thebibliography}
\end{document}